\documentclass[]{aa}       
\usepackage{graphicx}
\usepackage{amssymb}
\usepackage{amsmath}
\usepackage{psfig}
\usepackage{longtable}
\usepackage{supertabular}

\begin{document}

\title{The termination region of high-mass microquasar jets}

\author{
        V. Bosch-Ramon \inst{1} \and
        M. Perucho\inst{2} \and
	P. Bordas\inst{3,4}
    }

\institute{Dublin Institute for Advanced Studies, 31 Fitzwilliam Place, Dublin 2, Ireland; valenti@cp.dias.ie
\and
Dept. d'Astronomia i Astrof\'{\i}sica, Universitat de Val\`encia, C/ Dr. Moliner 50, 46100, Burjassot (Val\`encia), 
Spain; Manel.Perucho@uv.es 
\and
Institut f\"ur Astronomie und Astrophysik, Universit\"at T\"ubingen, Sand 1, 72076 T\"ubingen, Germany
\and
{\textit{INTEGRAL}} Science Data Centre, Universit\'e de Gen\`eve, Chemin d'Ecogia 16, CH--1290 Versoix, Switzerland; 
pol.bordas@uni-tuebingen.de}

\titlerunning{The termination region of HMMQ jets}

\offprints{V. Bosch-Ramon, \email{vbosch@mpi-hd.mpg.de}}

\date{Received <date> / Accepted <date>}

\abstract
{The environment of high-mass X-ray binaries can be characterized either by the SNR that forms these systems, or by
the wind from the companion massive star. These regions should be tenuous but very hot, and surrounded by a dense and cold 
shocked ISM shell. The interaction between the jet and such a complex medium, also affected by the system proper motion, 
can lead to very different jet termination structures.} 
{The evolution of the jet termination regions during the life of a high-mass microquasar is simulated to improve
the present understanding of these structures. Also, 
the evolving emission characteristics are modeled 
to inform potential
observational campaigns for this class of object.} 
{We have performed 2D numerical simulations of jets propagating in different scenarios, corresponding to 
different epochs after the formation of the high-mass
X-ray binary, using the code \textit{Ratpenat}. We have also made simple estimates of the non-thermal emission
that could be produced in the jet termination regions.}
{We find that, in the way through the hot and tenuous medium of the shocked wind/SNR ejecta, the jet suffers recollimation shocks in which
it loses part of its thrust and ends in a strong shock inflating a hot cocoon. 
The jet head propagates with a speed similar to the medium
sound speed, until it eventually reaches the denser and colder shocked ISM and the unperturbed ISM later on. In these last stages of evolution, the jet
is significantly slowed down and can be disrupted. For relatively old 
sources, the microquasar peculiar velocity becomes important, leading to complete jet destruction. 
Extended non-thermal radiation can be generated in the jet termination regions, being hard X-rays and TeV photons 
the best suited wavelengths to observe these structures.}{}

\maketitle

\section{Introduction} \label{intro}

High-mass X-ray binaries (HMXB) are the parent population of high-mass microquasars (HMMQ), the latter having the
peculiarity of presenting radio jet activity (e.g. Mirabel \& Rodr\'igues \cite{mirabel99}; Rib\'o \cite{ribo05}).
Jets of HMMQs propagate in a medium characterized by the stellar wind of the massive primary star of the system. At
the scales of the binary, the stellar wind can significantly affect the dynamics of the jet, trigger shocks suitable
for particle acceleration, and under some conditions, disrupt the jet by a combination of asymmetric recollimation
shocks and hydrodynamical instabilities (see Perucho \& Bosch-Ramon 2008 and Perucho et al. 2010). 

When launched, powerful enough jets can traverse the companion wind, and once at distances from the injection
point much larger than the separation distance ($z\gg d_{\rm orb}$, where $z$ is the axial coordinate and distance to
the compact object), they do not suffer significant interference by the supersonic wind, which comes from behind the
jet. At this stage, the jet head propagates with a speed $\sim \sqrt{n_{\rm j}(d_{\rm orb})/n_{\rm w}(d_{\rm
orb})}\,v_{\rm j}\sim 0.1\,v_{\rm j}$ assuming free expansion, where $v_{\rm j}$ is the initial jet velocity. In this
propagation phase, a strong shock will form at the tip of the jet, and particle acceleration may tap a significant
fraction of the jet luminosity there.

When the stellar wind is dynamically negligible, the jet can propagate as far and fast as the inertia of the swept ISM permits
and, eventually, the jet head is shocked inflating a wide cocoon that pushes a shell of shocked ISM (Bordas et al. 2009), as in
FRII galaxies (e.g. Fanaroff \& Riley 1974; Kaiser \& Alexander 1997). However, for systems hosting OB stars, the dynamics
of the jet may be severely affected before reaching the unperturbed ISM. 

In order to better understand the hydrodynamical jet evolution at different stages of a HMXB, we have performed 2D
simulations of HMMQ jets for three different scenarios: soon after the SNR explosion (case 1), the phase of wind/ISM
interaction (case 2), and the phase in which the wind/ISM interaction is affected by the system proper motion (case 3). We have
also estimated the broadband non-thermal emission that may be produced in cases 1 and 2, and made predictions of their main
observational features.

\section{The physical picture} \label{intro}

For a young HMXB, say of age $t_{\rm src}\sim 3\times 10^4$~yr, with typical ISM densities $n_{\rm ISM}\sim
1$~cm$^{-1}$ and moderate peculiar velocities $v_{\rm p}\la 10^7$~cm~s$^{-1}$, the system may still be embedded inside
an adiabatically expanding SNR (case 1). In such a situation, if a jet is launched, after $\sim 10$~yr it will
encounter a region of shocked stellar wind confined by the shocked SNR ejecta, the latter to its turn driving a strong
shock in the ISM. Once in the high-pressure region of the shocked stellar wind, the jet head will be slowed down to
medium subsonic velocities ($\sim 1000$~km~s$^{-1}$ for typical jet and wind properties, see below). At farther
distances, the jet head will encounter the shocked SNR ejecta eventually reaching the shocked ISM. Finally, the jet
will enter the normal ISM having lost significant amounts of thrust, producing a shock in the ISM and possibly
suffering disruption. Intermittent jet activity with low duty cycle will not clean the way for later ejections, since
the sound speed in the shocked wind and the SNR ejecta is high. High enough duty cycles ($\ga 10$\%) for the jet to
keep clean its way could be understood in terms of an averaged persistent activity. The case of a HMMQ persistent jet
interacting with a SNR has a clear example in SS~433 (see Vel\'azquez \& Raga 2000, Zavala et al. 2008).

Older HMXBs have already left the SNR, and the stellar winds can directly interact with the ISM (case 2). In
this case, the stellar wind inflates a bubble filled with shocked wind material, and this hot bubble drives a shock in
the ISM that becomes radiative even for moderate ages of the source ($t_{\rm src}\sim 10^5$~yr; see Castor et al.
1975). Therefore, a jet launched in this HMXB phase will interact first with the shocked wind bubble, reaching later
on a thin, dense and cold shell of shocked ISM. Qualitatively, the situation is similar to the previous case, but now
the ISM shell is much denser because of radiative cooling. Eventually, the jet can reach the unperturbed ISM, shocking
it. As in case 1, the jet may be eventually disrupted. Cygnus~X-1 could correspond to the situation represented by case 2 (Mart\'i et al.
1996; Gallo et~al. 2005, Russell et al. 2007). Since there are
not clear signatures of a reverse shock in this source (Bordas et al. 2009), and only the shocked ISM thermal
radiation is detected, it may be that the Cygnus~X-1 jet head is fully disrupted. A shock driven in the ISM can still
develop in such a conditions since the disrupted jet region has much higher pressure than the environment as
simulations of FRI radio galaxies show (Perucho \& Mart\'i 2007; see also Bordas et al. 2010a).

If the HMXB age is old enough, say $t_{\rm src}\sim 10^6$~yr, the wind-driven ISM shock can be slower than $v_{\rm
p}$. Then, a bow-shaped shock forms, driven by the system proper motion and pointing in its direction (case 3). 
In this scenario, the shocked ISM material is slowly evacuated in the opposite direction to the system motion, and the shocked wind
material is also advected in the same direction but at much faster velocities, similar to shocked wind sound speed
($\sim 10^3$~km~s$^{-1}$). If the jet is launched in this stage, for jet directions misaligned enough with the binary
motion the shocked wind material will impact the jet roughly from the side. At farther distances, the jet will
encounter the shocked ISM, roughly as dense and cold as in case 2. The dense shocked ISM layer plus the shocked wind
lateral impact will stop the jet advance and lead to total jet destruction before reaching the unshocked ISM.

\begin{figure}[!h]
   \centering
\includegraphics[clip,angle=0,width=0.7\columnwidth]{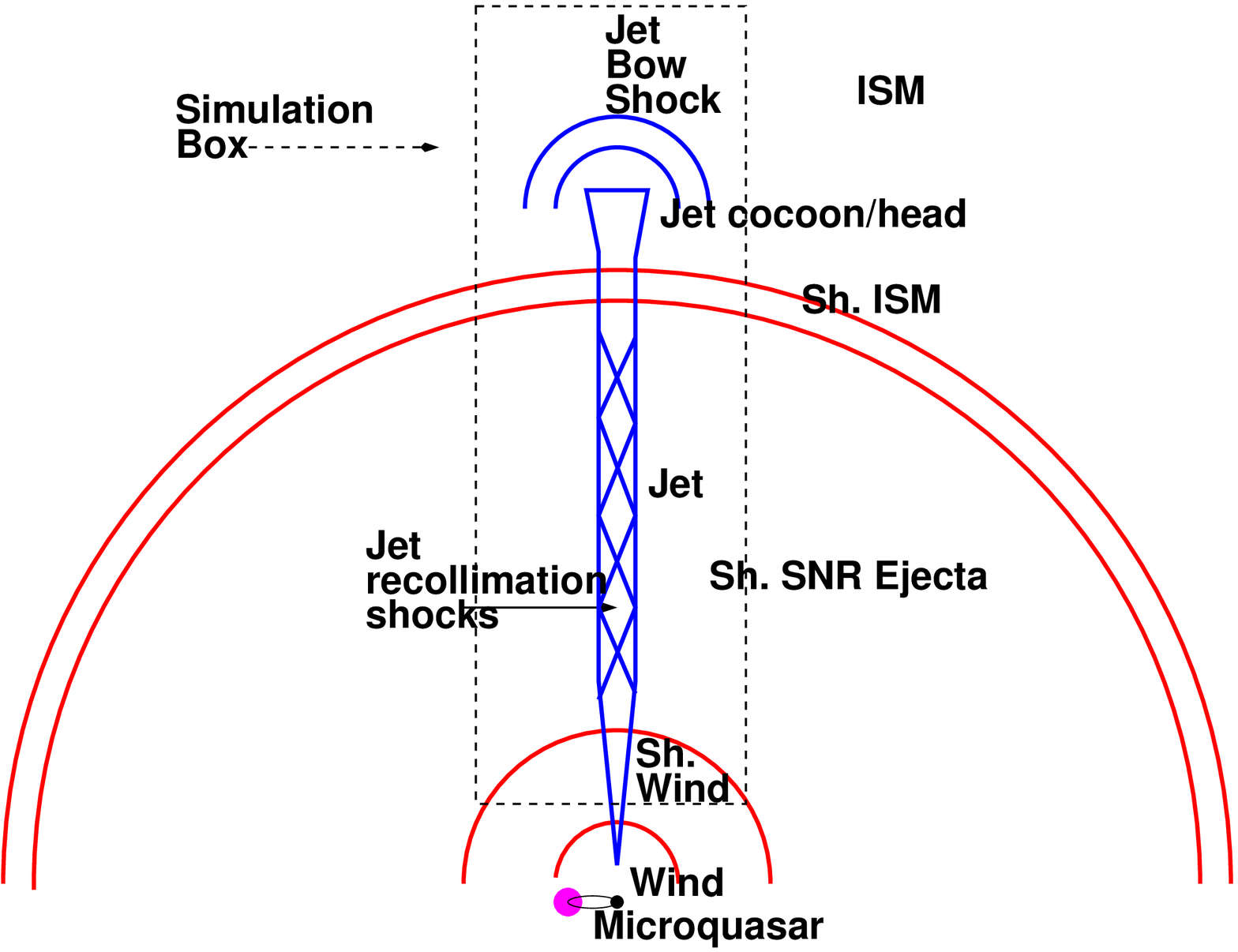}
\includegraphics[clip,angle=0,width=0.7\columnwidth]{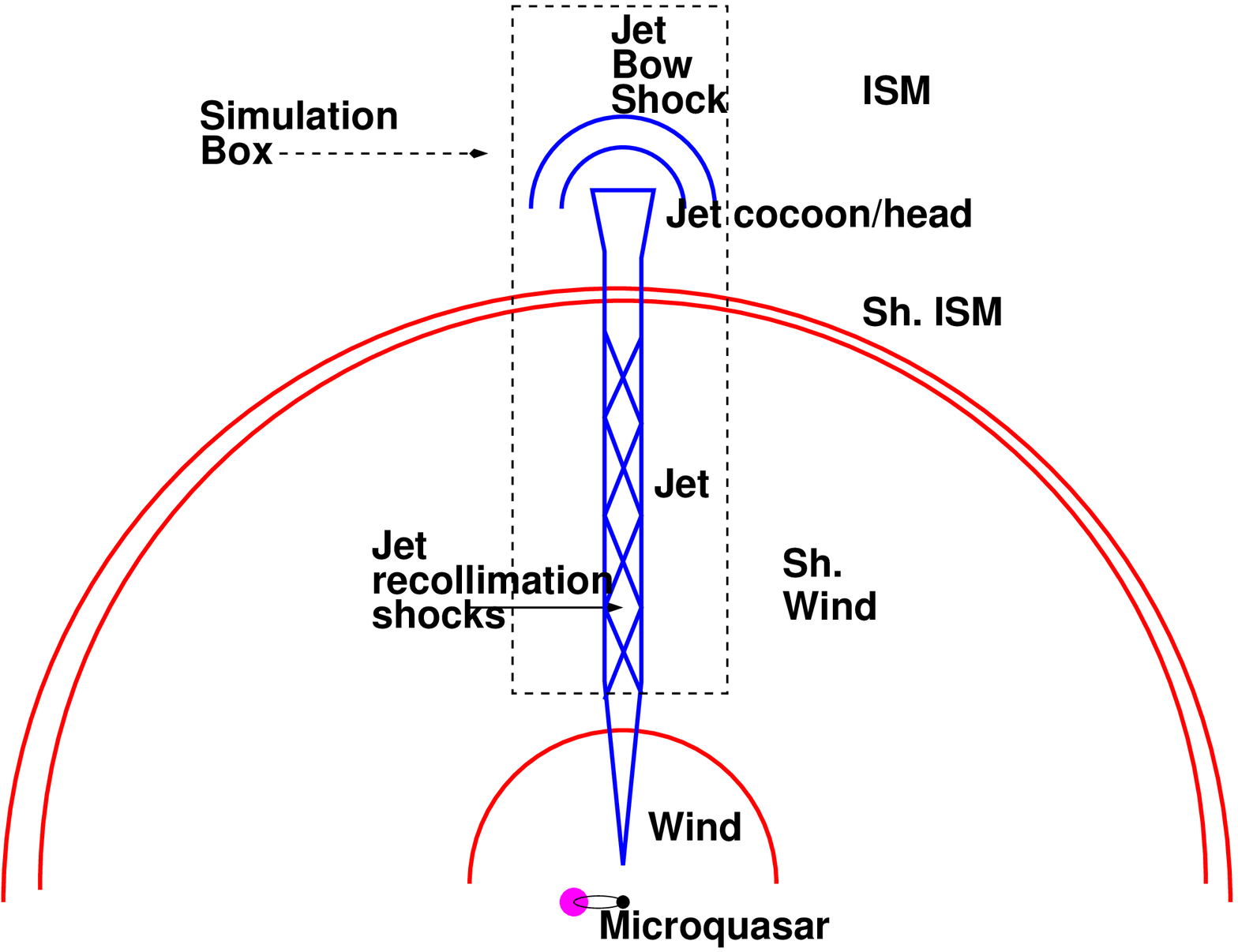}
\includegraphics[clip,angle=0,width=0.7\columnwidth]{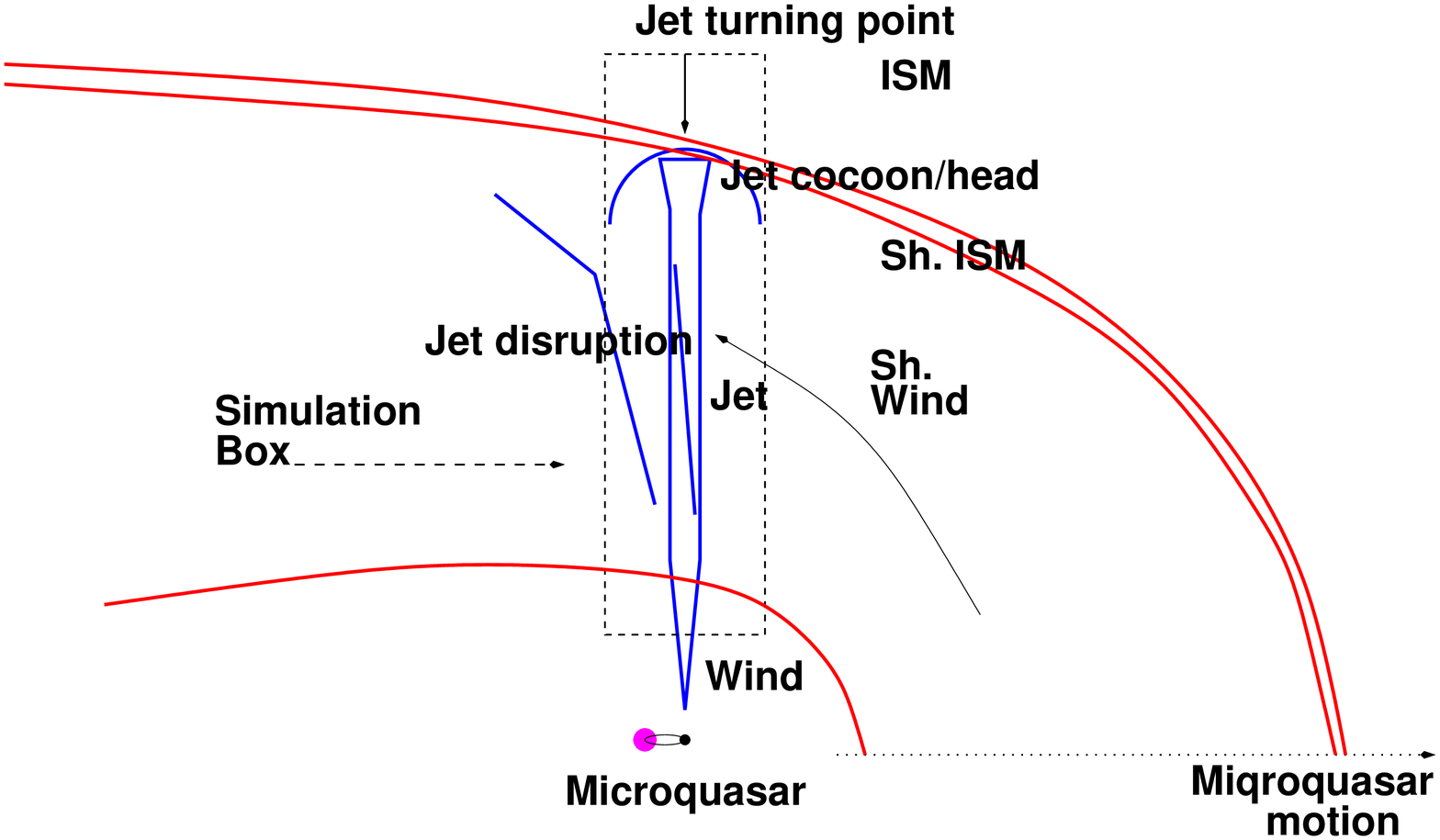}
\caption{Sketches of a 
jet propagating through: the shocked stellar wind, the shocked
SNR ejecta, and the shocked and unshocked ISM (top); 
the shocked stellar wind, and the shocked and unshocked ISM (middle); 
through the unshocked and shocked stellar wind, 
and interacting with the shocked ISM under a significant peculiar velocity (bottom).}
\label{casos}
\end{figure}

Sketches of the scenarios corresponding to cases 1, 2 and 3 are presented in the top, middle and bottom panels of Fig.~\ref{casos}.

\section{Studying the jet termination}

\subsection{Medium characteristics}\label{med}

We have adopted an analytical approach to characterize the medium in which the jet propagates. In the three cases
considered here, the ISM density has been fixed to $n_{\rm ISM}=1$~cm$^{-3}$, with a temperature $T_{\rm ISM}=10^4$~K,
as expected in regions close to massive stars. The stellar wind velocity and the stellar mass-loss rate have been
taken to be $v_{\rm w}=2\times 10^8$~cm~s$^{-1}$ and $\dot{M}_{\rm w}=10^{-6}\,M_\odot$/yr, respectively (typical for
an O star; see, e.g., Puls et al. 2009), which corresponds to a wind kinetic luminosity $L_{\rm w}=\dot{M}_{\rm
w}\,v_{\rm w}^2/2\approx 1.3\times 10^{36}$~erg~s$^{-1}$. 

In case 1, the system is assumed to be at the center of a SNR. The ejected mass of the SNR has been taken to be
2~$M_\odot$, and the explosion energy $5\times 10^{49}$~erg. These moderate values have been adopted accounting for
the fact that the energetics of supernovae leading to the formation of some HMXBs could be quite low (a relatively extreme
case may be Cygnus~X-1; see Mirabel \& Rodrigues 2003). We have calculated the properties of the shocked ejecta of
the SNR and the shocked ISM adopting the solution of a spherical explosion in its adiabatic phase (e.g. Blandford \&
McKee 1976) for $t_{\rm src}\sim 3\times 10^4$~yr. The radius of the whole region affected by the SNR 
is
$z_{\rm bs}\approx (E_{\rm SNR}/m_{\rm H}n_{\rm ISM})^{1/5}\,t_{\rm src}^{2/5}\approx 3\times
10^{19}$~cm for the adopted parameter values, where $m_{\rm H}$ is the hydrogen mass. The density of the shocked ISM
is the one for a strong adiabatic shock, $n_{\rm ISMs}=4\,n_{\rm ISM}$, and the temperature, $T_{\rm ISMs}\approx
4\times 10^5$~K. From $z_{\rm bs}$ and $E_{\rm SNR}$, we derive a characteristic pressure within that radius,
$P_{\rm SNR}\sim 2\,E_{\rm SNR}/3\,V_{\rm bs}\approx 3\times 10^{-10}$~erg~cm$^{-3}$ ($T_{\rm SNR}\approx 10^8$~K; $v_{\rm
bs}=4\,\pi\,z_{\rm bs}^3/3$), as well as the locations of the contact discontinuity between the shocked ISM and the
shocked ejecta, $z_{\rm c}\approx 0.9\,z_{\rm bs}$. The location of the contact discontinuity between the shocked
stellar wind and the SNR ejecta is computed from taking the same pressure in both sides, obtaining $z_{\rm
ws}\approx 8\times 10^{18}$~cm. The shocked wind density is $n_{\rm ws}\approx 0.02$~m$^{-3}$, and the
temperature $T_{\rm ws}\approx 10^8$~K. We simulate the jet once it is already inside the shocked wind region, since at
closer distances the dynamical impact of the medium on the jet is negligible.

In case 2, we have adopted the solution given by Castor et al. (1975) that describes the interaction of a stellar wind
with the ISM at the stage when the shock in the latter is already radiative. We have adopted in this case an age of
$10^5$~yr, and the radius of the whole interaction region is $z_{\rm bs}=0.76\,(L_{\rm w}/\,m_{\rm H}\,n_{\rm
ISM})^{1/5}\,t^{3/5}\approx 2\times 10^{19}$~cm. The pressure in the shocked wind region is $P_{\rm ws}\approx 
n_{\rm ISM}\,m_{\rm H}\,\dot{z}_{\rm bs}^2\approx 3\times 10^{-11}$~cm~s$^{-1}$ ($T_{\rm ws}\approx 6\times 10^7$~K), and the density, $n_{\rm
ws}=4\times 10^{-3}$~cm$^{-3}$.  The shocked ISM region is a very thin, cool and dense layer. For simplicity, the
width of this region has been taken equal to one simulation cell: $10^{17}$~cm, with $n_{\rm ISMs}\approx
10^2$~cm$^{-3}$ and $T_{\rm ISMs}=2\times 10^3$~K. As in case 1, the jet injection point is already in the wind
shocked region.

In case 3, taking $v_{\rm p}=10^7$~cm~s$^{-1}$ and $t_{\rm src}=10^6$~yr, one gets $\dot{z}_{\rm bs}\sim 10^6\,{\rm
cm~s}^{-1}\ll v_{\rm p}$ and the wind-driven shock in the ISM becomes strongly asymmetric, with the shocked wind
material advected backwards with respect to this shock. The wind driven shock in the ISM is now closer to the HMMQ
than in case 2, at $z_{\rm w}\approx 8\times 10^{18}$~cm, at which the (pre-shock) stellar wind density and
temperature are $n_{\rm w}=2.4\times 10^{-4}$~cm$^{-3}$ and $T_{\rm w}=10^3$~K. The density, pressure and temperature
in the shocked wind are $n_{\rm ws}=10^{-3}$~cm$^{-3}$, $P_{\rm ws}=1.4\times 10^{-11}$~erg~cm$^{-3}$ and $T_{\rm
ws}=10^8$~K, respectively, and a lateral shocked wind velocity $v_{\rm ws\perp}=10^8$~cm~s$^{-1}$. The
shocked wind region reaches $z_{\rm ws}=10^{19}$~cm, in which a thin shocked ISM shell, of $10^{17}$~cm width, $n_{\rm
ISMs}=100$~cm$^{-3}$ and $T_{\rm ISMs}=10^3$~K, is located. In this simulation, the jet is launched in the unshocked
wind region, because it is more extended than in previous cases, and the unshocked wind is taken to come from behind
the jet. 

The values of the main ambient medium parameters, for the three cases considered here, are summarized in
Table~1.

\begin{table}{
\begin{center} 
  \begin{tabular}{lccc}
    \hline
    Case 1 & Jet height  & medium dens.  & medium temp. \\
     & $z$ (cm)& $n$ ($m_p$~cm${}^{-3}$) & $T$ (K)   \\
     \hline
    Shocked  wind& $3\times 10^{18}$ &  0.02 & $10^8$  \\
    SNR& $8\times10^{18}$ & 0.02 & $10^8$\\
    Shocked ISM&  $3\times10^{19}$ & 4.0 & $4\times 10^5$ \\
    ISM &  $3.3\times10^{19}$& 1.0 & $10^4$\\
    \hline
  Case 2   &   &  &  \\
    \hline
    Shocked  wind& $3\times 10^{18}$ &  $4\times10^{-3}$ & $6\times10^7$  \\
    Shocked ISM&  $2\times10^{19}$ & 100 & $2\times 10^3$ \\
    ISM &  $2.01\times10^{19}$& 1.0 & $10^4$\\
    \hline
  Case 3   &  & &   \\
    \hline
    Stellar  wind& $1.25\times 10^{18}$ & $2.4\times10^{-4}$  & $10^3$  \\
    Shocked wind& $8\times10^{18}$ & $10^{-3}$ & $10^8$ \\
    Shocked ISM&  $1\times10^{19}$ & 100 & $10^3$  \\
    ISM &  $1.01\times10^{19}$& 1.0 & $10^4$  \\
  \end{tabular}
\end{center}
\caption{The ambient medium in cases 1, 2 and 3.}
}
\end{table}

\subsection{Simulations}

We have performed 2D simulations of jets evolving in the three different scenarios described above. In cases 1 and 2
the jets are simulated using an axisymmetric grid in cylindrical coordinates, whereas the transversal motion of the
shocked wind in case 3 requires a slab jet in planar coordinates. We expect from our previous experience (see Perucho
\& Bosch-Ramon 2008, Perucho et al. 2010) that the results of the slab simulation will be qualitatively correct. The
jets are injected with power $L_{\rm j}=3\times 10^{36}$~erg~s$^{-1}$ at $z_0=3\times 10^{18}\,\rm{cm}$ for cases 1
and 2, and $z_0=1.25\times10^{18}\,\rm{cm}$ for case 3. In case 3, the power corresponds to a jet square section (i.e.
the power crossing a surface equal to the jet width square). In the cases 1 and 2, the jet is assumed to have a radius at the
injection point $R_{\rm j0}=0.1\,z_0$. In case 3, the jet half-side length has been taken as $1.5\times 10^{17}$~cm. 
We have fixed the
jet velocity at $v_{\rm j}=10^{10}$~cm~s${}^{-1}$, the Mach number, $M_{\rm j}\approx
18$, and the jet temperature, $T_{\rm j}=1.2\times10^9\,\rm{K}$ at injection in all three cases. The jets are composed by protons and electrons and
have densities of $\rho_{\rm j}=1.8\times 10^{-29}$~g~cm$^{-3}$ for cases 1 and 2, and $\rho_{\rm j}=8.3\times
10^{-29}$~g~cm$^{-3}$ for case 3. The adiabatic exponent for the jets, with the given temperature and composition is
$\Gamma\approx 1.58$. The ambient medium in the different simulations follows the description given in
Sect.~\ref{med}. We note that the shock driven by the jet head in the medium (wind, SNR and ISM) is called here bow
shock (which is more a sound wave rather than a real shock in the hot shocked wind and SNR ejecta), and the jet
shocked region is called cocoon, following Bordas et al. (2009).

The simulations have been carried out with the finite-volume code \textit{Ratpenat}, which solves the
equations of relativistic hydrodynamics in  conservation form using high-resolution-shock-capturing methods.
\textit{Ratpenat} was parallelized with a hybrid scheme with both parallel processes (MPI) and parallel threads
(OpenMP) inside each process (see Perucho et al. 2010). The simulations were performed using the computational facility
{\it Tirant}, at the {\it ``Servei
d'Inform\`atica"} of the {\it ``Universitat de Val\`encia"}, with 12, 8 and 4 processors for cases 1, 2 and 3, respectively.
In the axisymmetric simulations, the numerical grid box expands transversally to 100~$R_{\rm j0}$, with an extended
grid of 100~$R_{\rm j0}$ added to the uniform grid. In the axial direction, the grid extends to $150\,R_{\rm j0}$ in
case 1, and to $100\,R_{\rm j0}$ in case 2. The numerical resolution of these simulation is of eight cells per
initial jet radius. The computational grid, including the extended region is $1200\times 1000$ cells. In the case of
the simulation of the slab jet, the grid expands $100\,R_{\rm j0}$ to each side of the axis, plus $100\,R_{\rm j0}$
extra at each side, and $100\,R_{\rm j0}$ in the axial direction. The resolution used here is of four cells per jet
initial half-side length.  The dimensions of the grid are thus $1200\times 400$ cells. The boundary conditions
adopted in the Cartesian and cylindrical cases are reflection at the jet bottom side, injection right at the jet
base, and outflow in the rest. Reflection in the jet axis has been also adopted in the cylindrical simulations.

\subsection{Results}

   \begin{figure}[!h]
     \centering
   \includegraphics[clip,angle=0,width=\columnwidth]{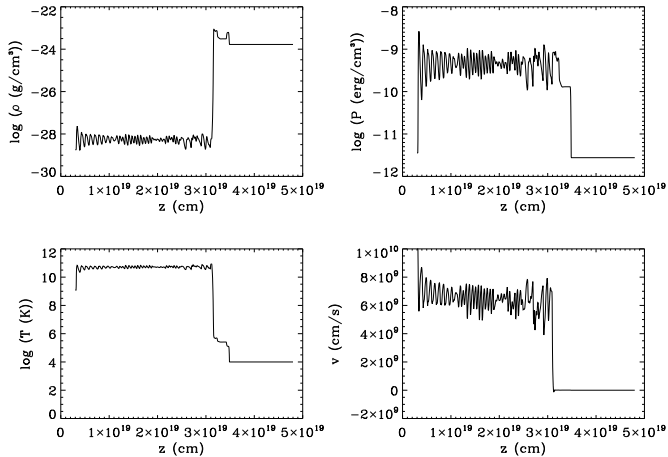}
   \caption{Axial cuts of density (top, left), pressure (top, right), 
   temperature (down, left) and velocity (down, right) 
   at the last snapshot of the evolution in case 1 ($t_{\rm src}=8900$~yr). Units are cgs.}
   \label{fig:ax1}
   \end{figure}

    \begin{figure}[!h]
     \centering
   \includegraphics[clip,angle=0,width=\columnwidth]{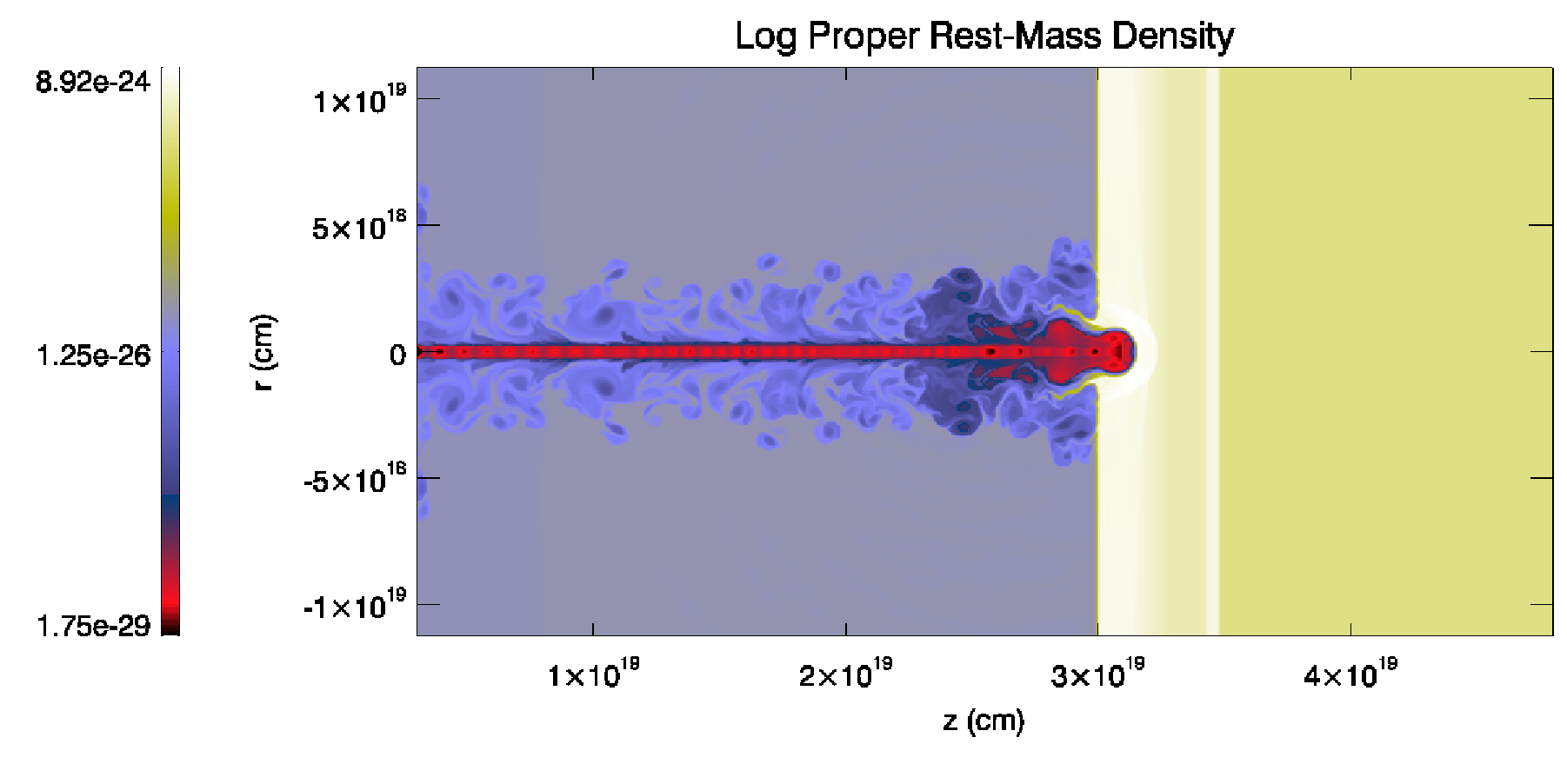}
   \includegraphics[clip,angle=0,width=\columnwidth]{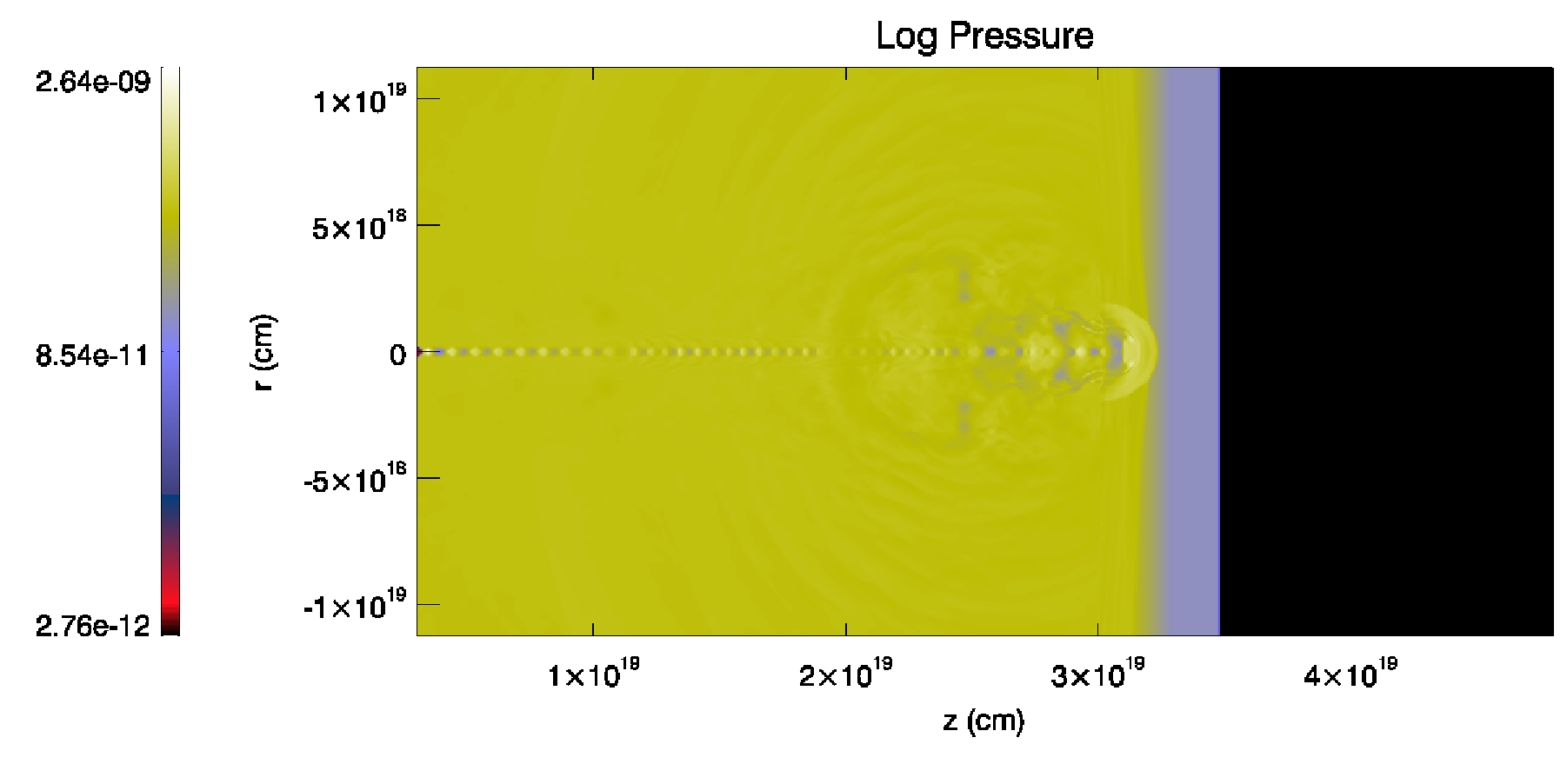}
   \includegraphics[clip,angle=0,width=\columnwidth]{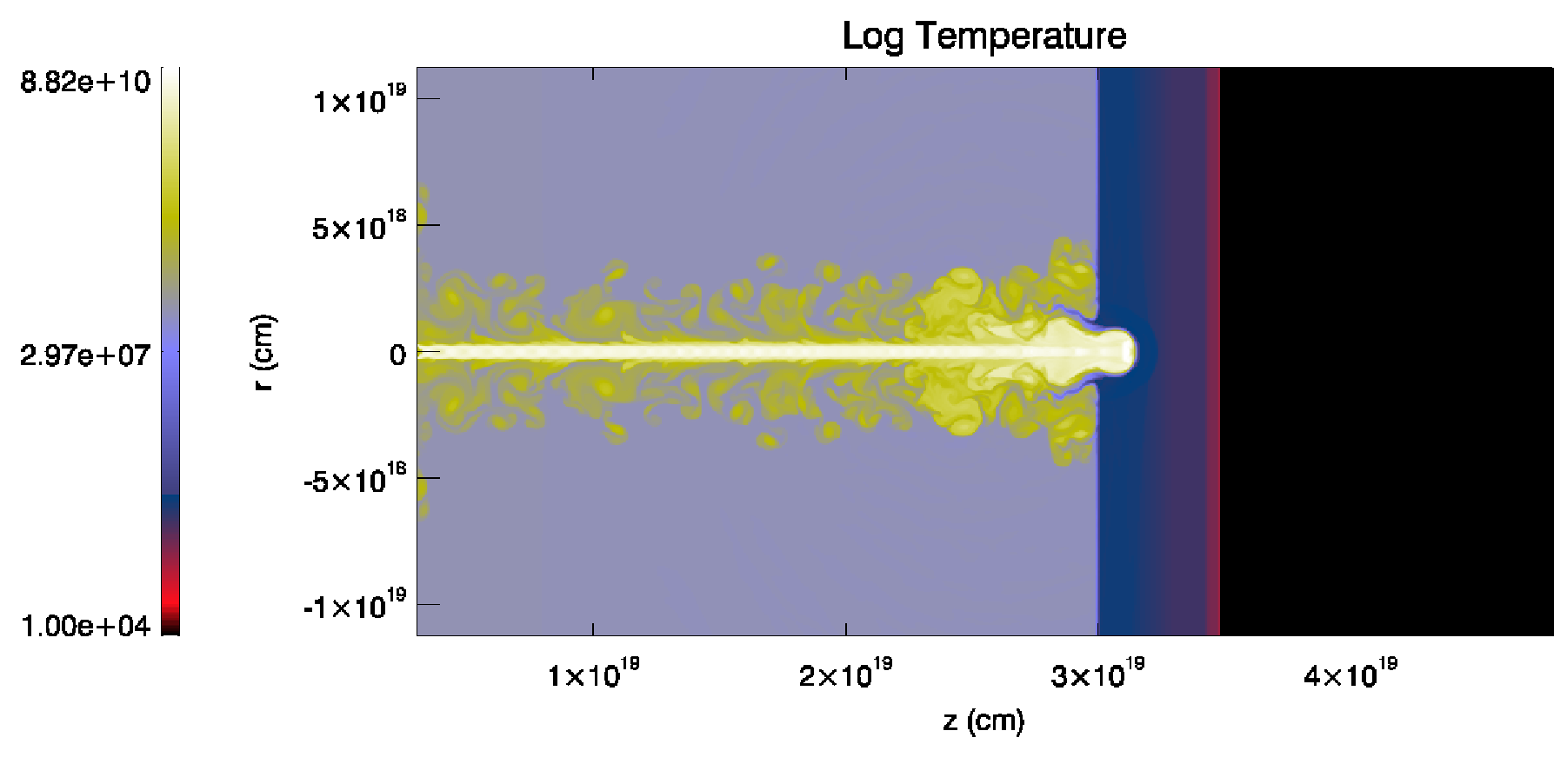}
   \includegraphics[clip,angle=0,width=\columnwidth]{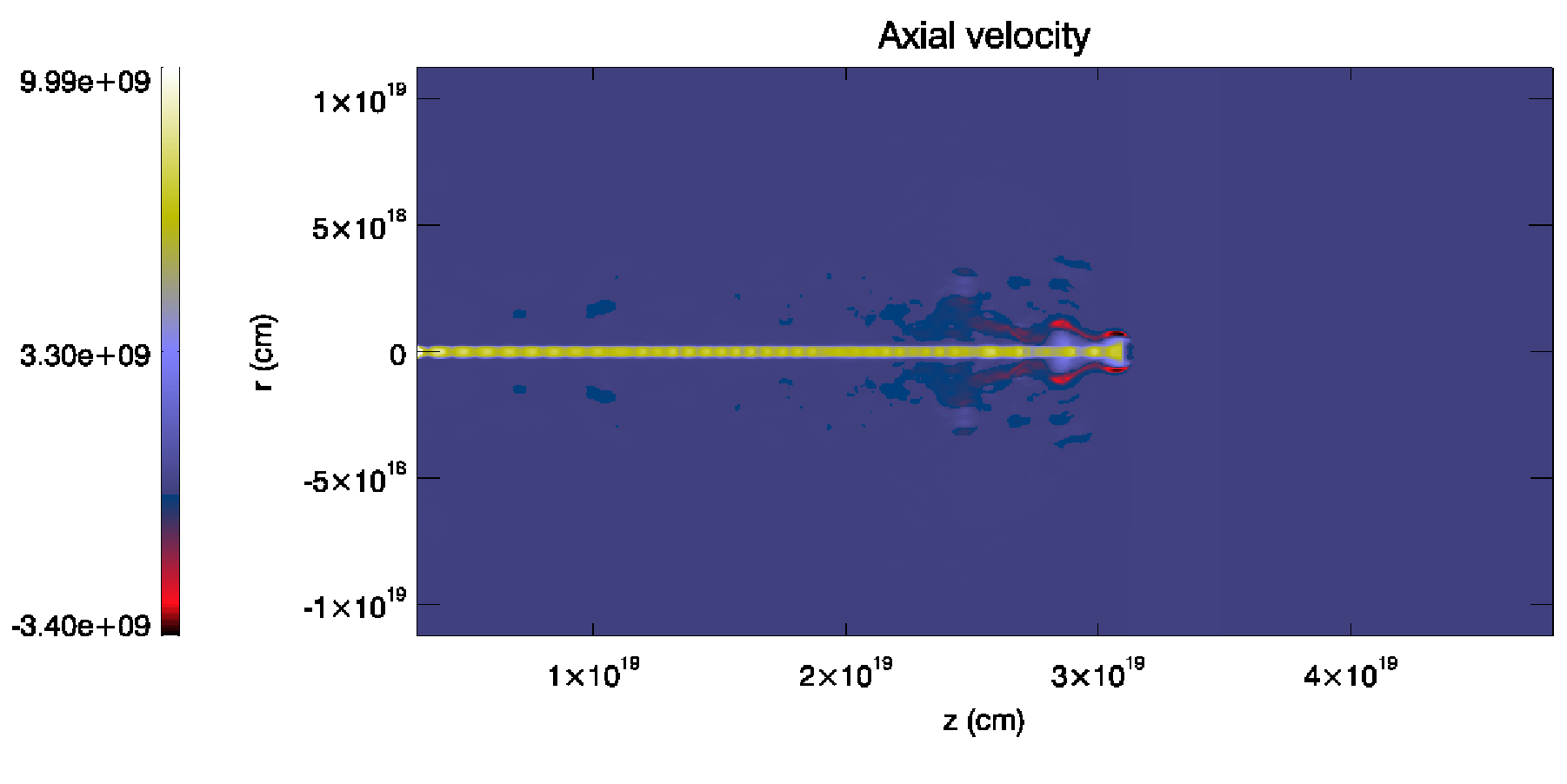}
   \caption{From top to bottom, 
   maps of density, pressure, temperature and axial velocity at the last snapshot of the evolution in case 1 ($t_{\rm src}=8900$~yr). Units are cgs.}
   \label{fig:maps1}
   \end{figure}

    \begin{figure}[!h]
     \centering
   \includegraphics[clip,angle=0,width=0.45\columnwidth]{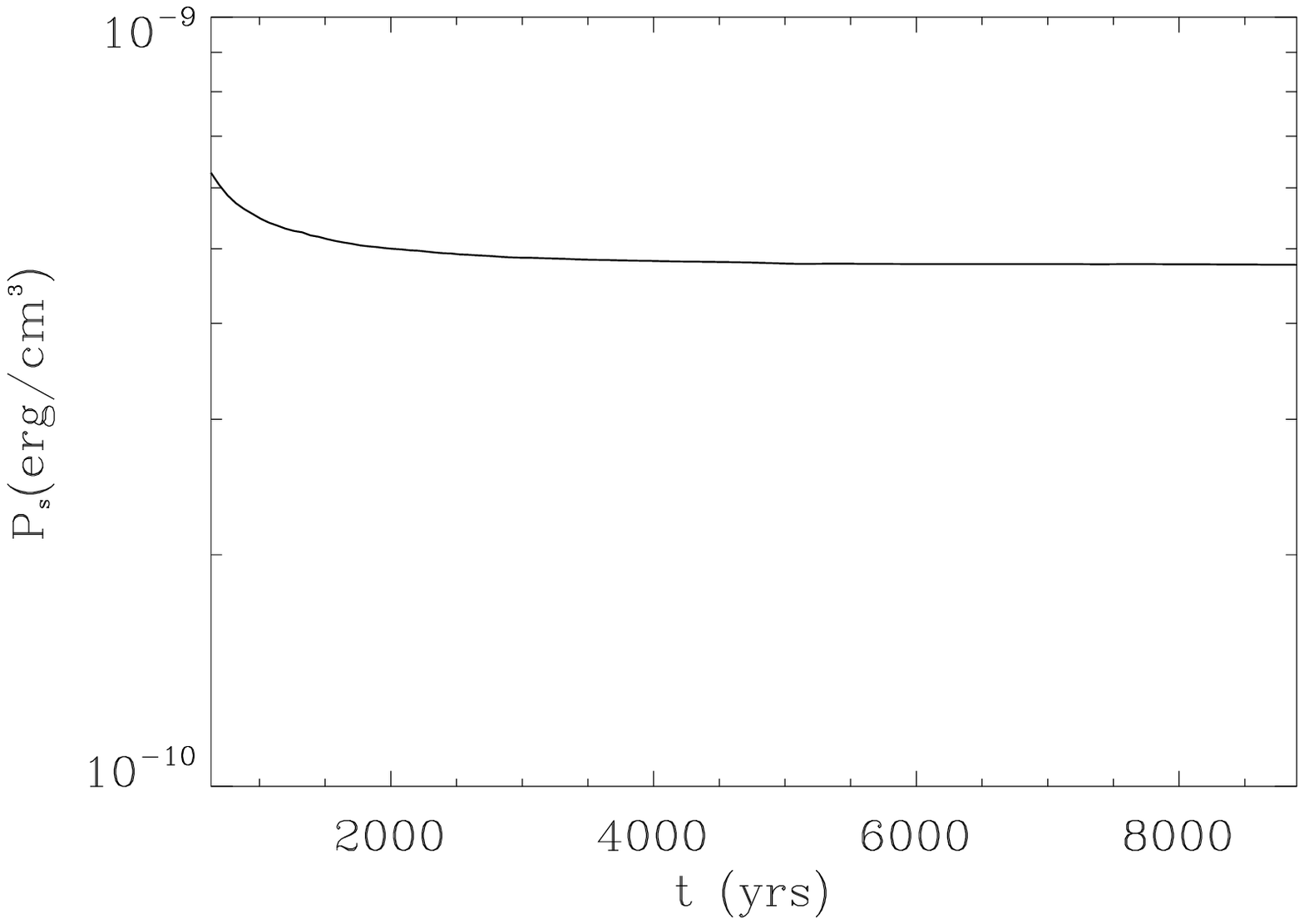}
   \includegraphics[clip,angle=0,width=0.45\columnwidth]{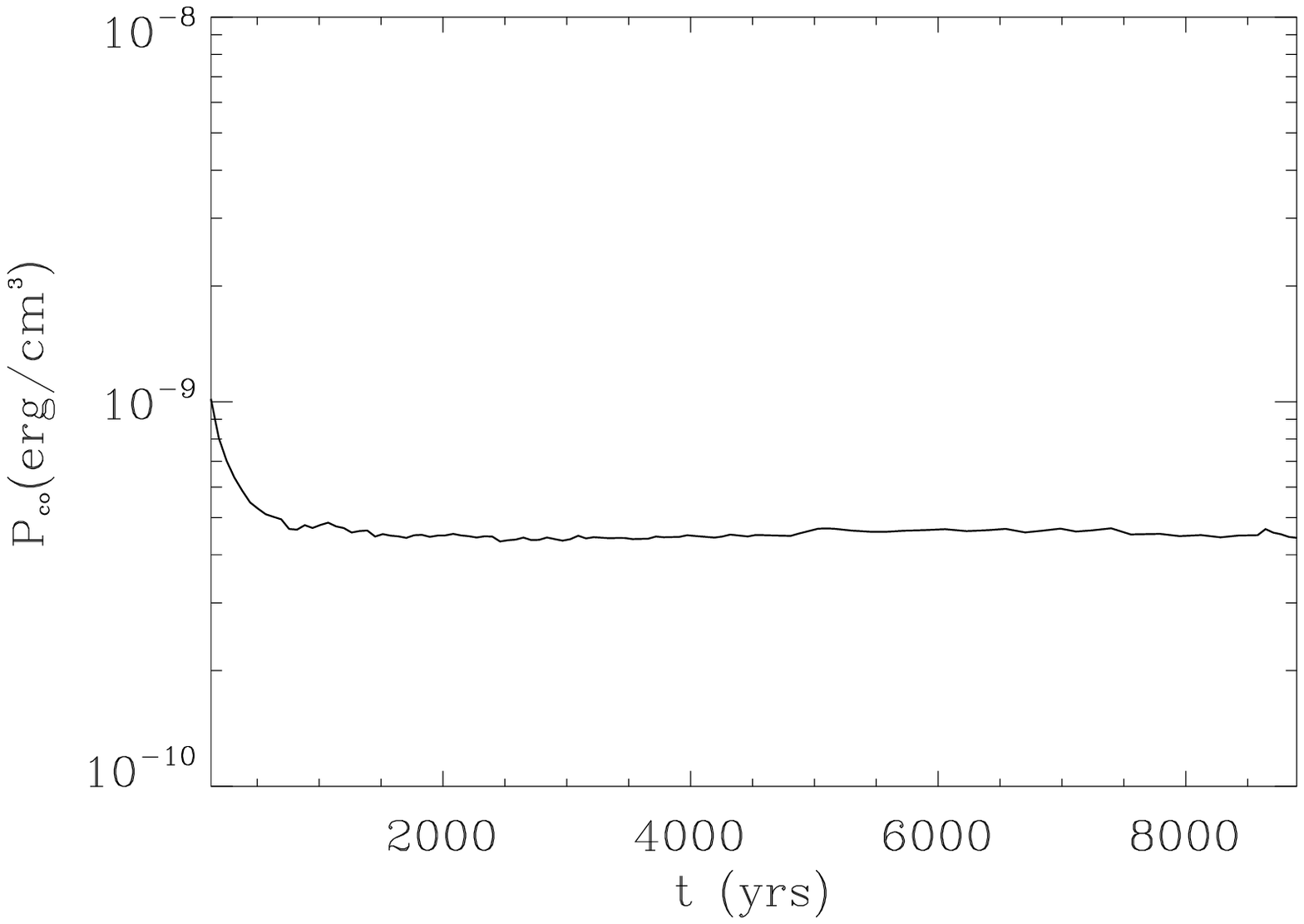}
   \includegraphics[clip,angle=0,width=0.45\columnwidth]{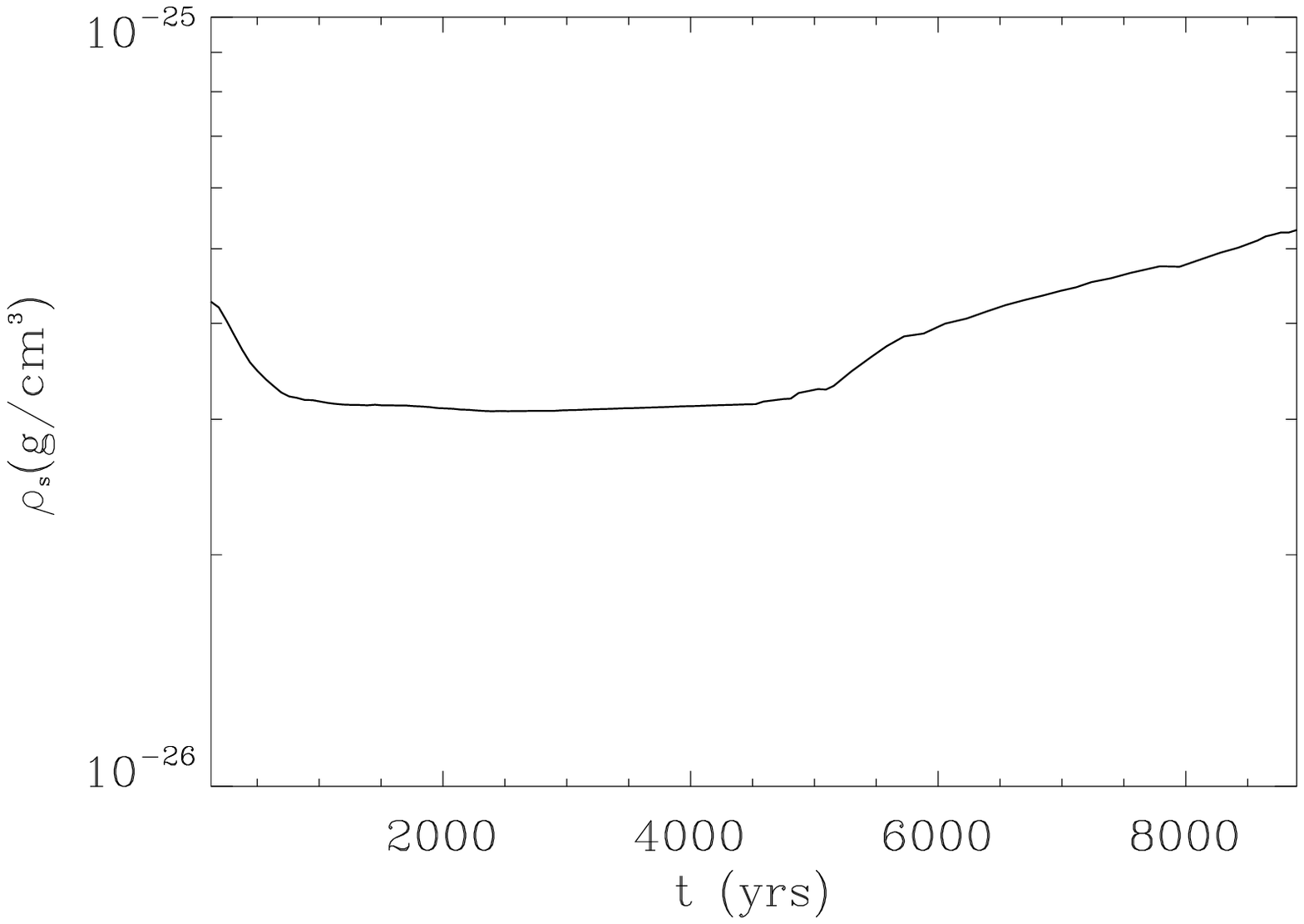}
   \includegraphics[clip,angle=0,width=0.45\columnwidth]{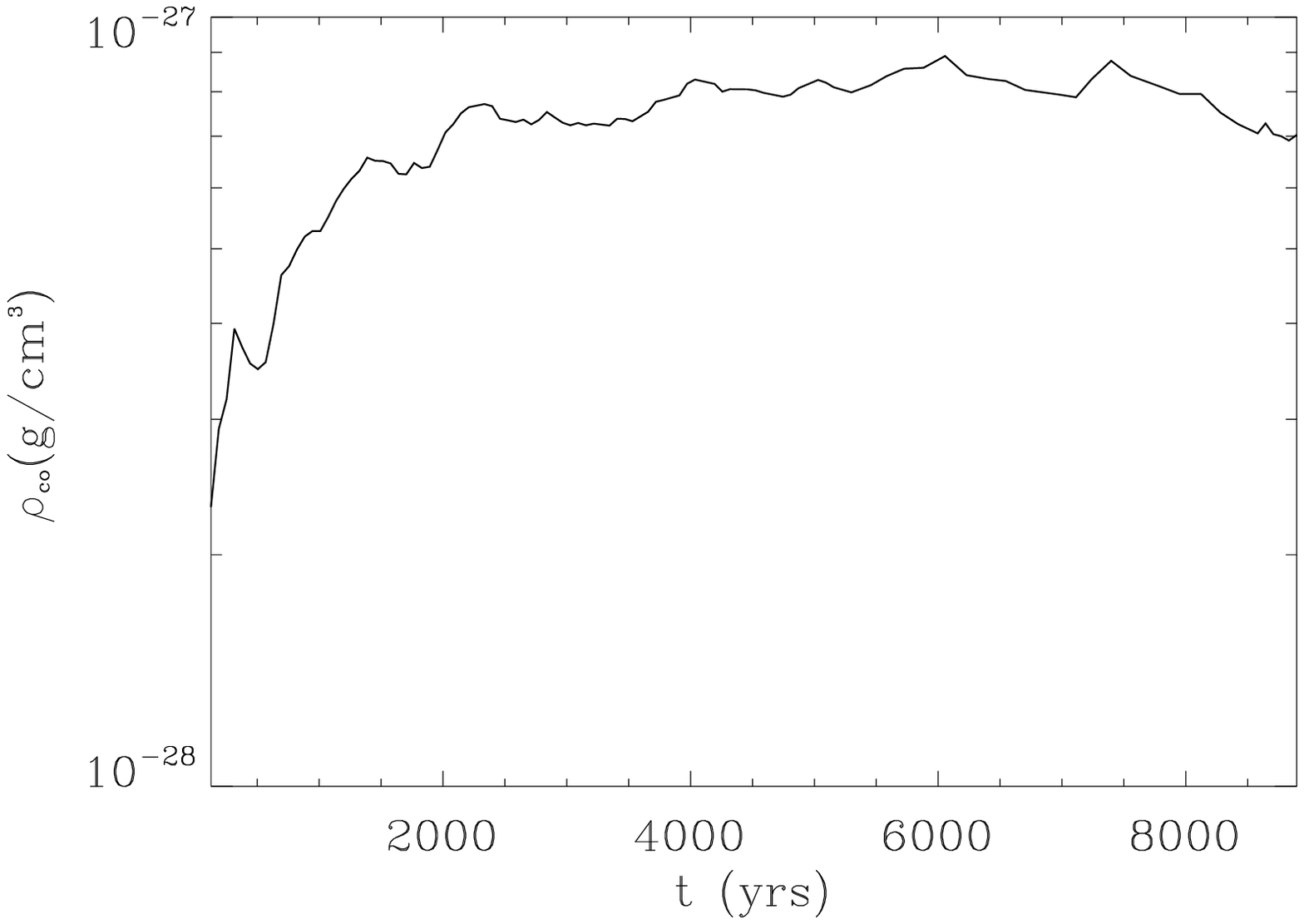}
   \includegraphics[clip,angle=0,width=0.45\columnwidth]{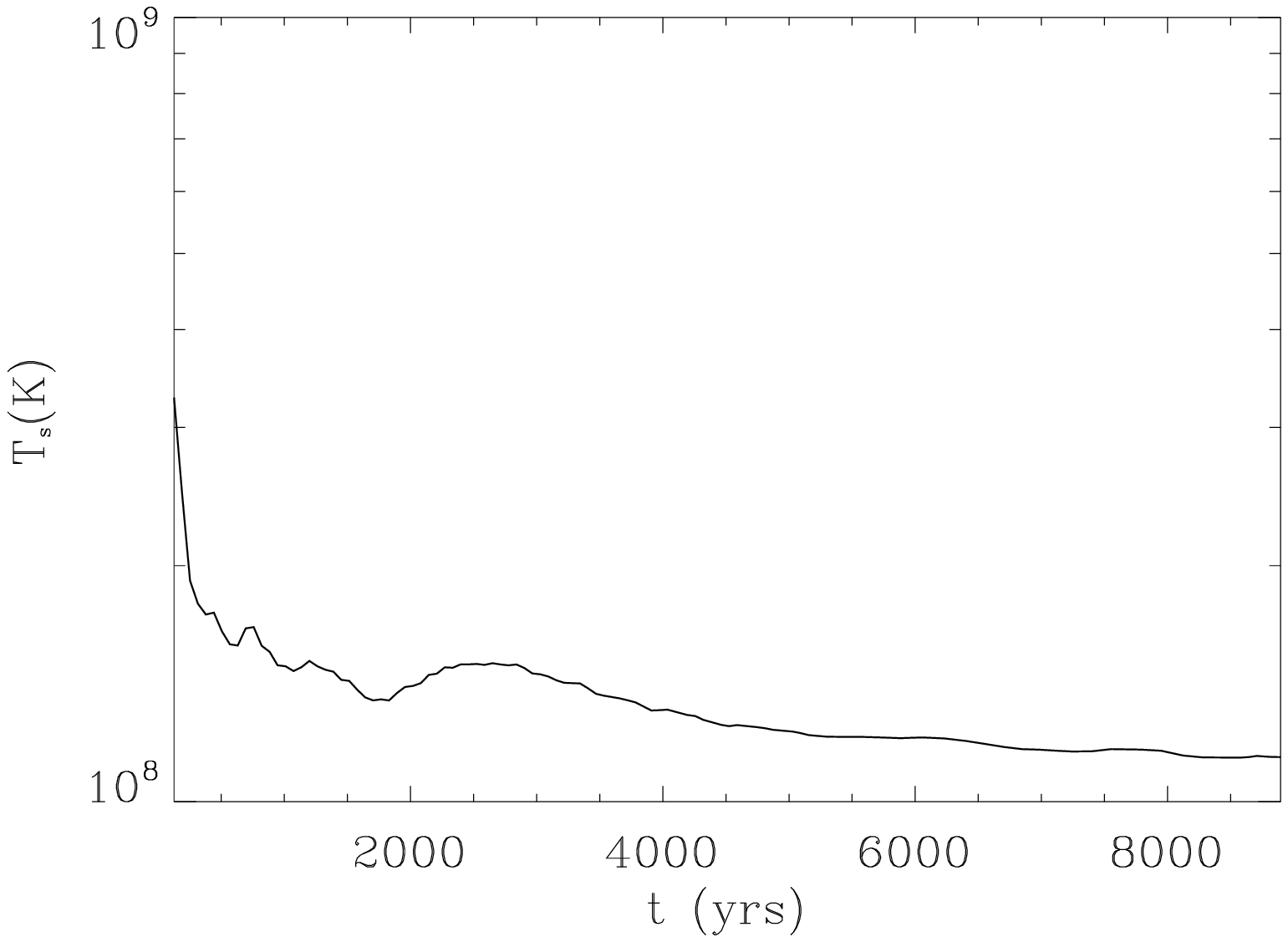}
   \includegraphics[clip,angle=0,width=0.45\columnwidth]{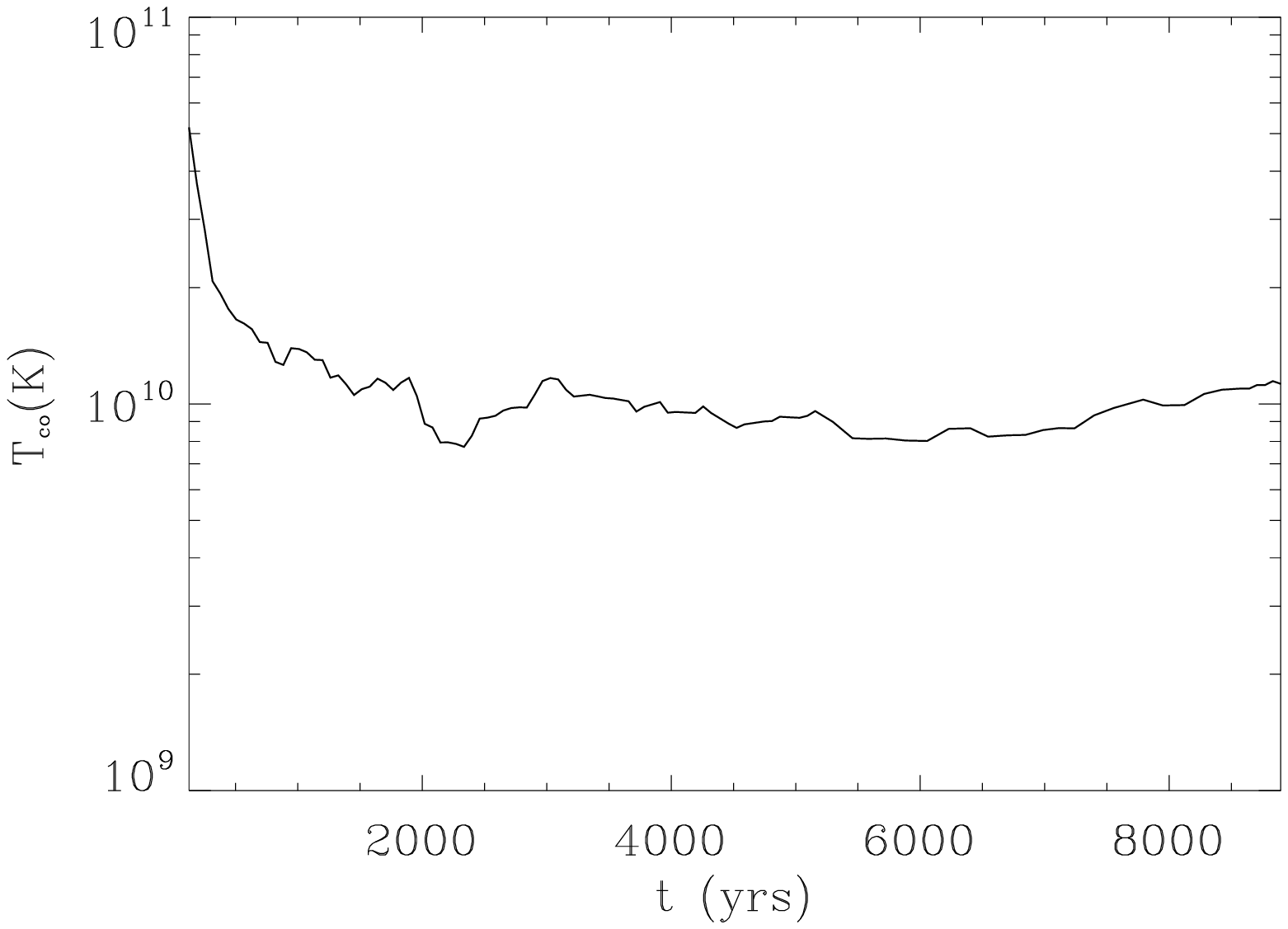}
   \caption{Evolution of pressure (top), density (middle), and temperature (bottom) in the bow shock (left) and cocoon (right) regions
   for case 1. Units are cgs.}
   \label{fig:evol12}
   \end{figure}

    \begin{figure}[!h]
     \centering
   \includegraphics[clip,angle=0,width=0.45\columnwidth]{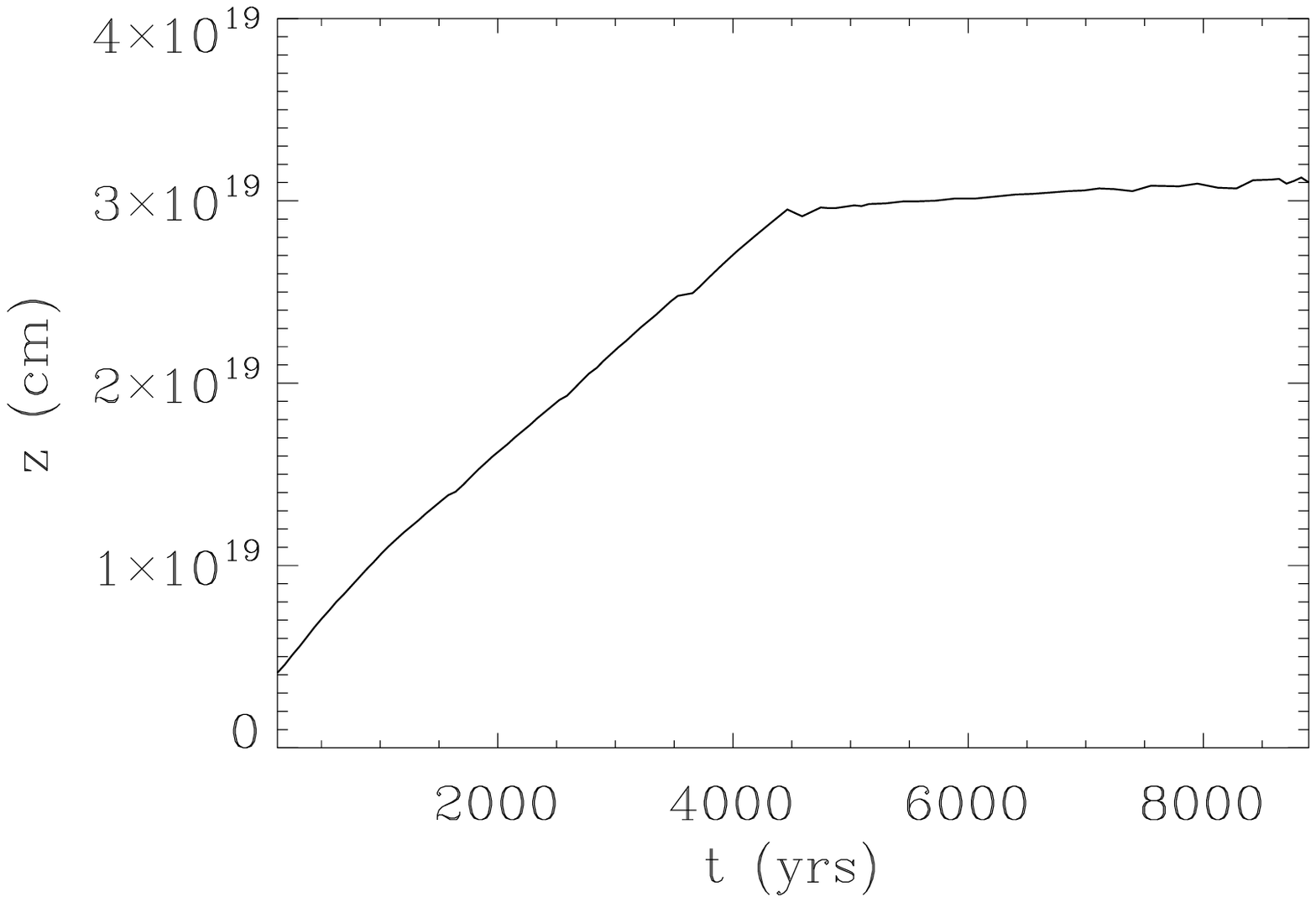}
   \includegraphics[clip,angle=0,width=0.45\columnwidth]{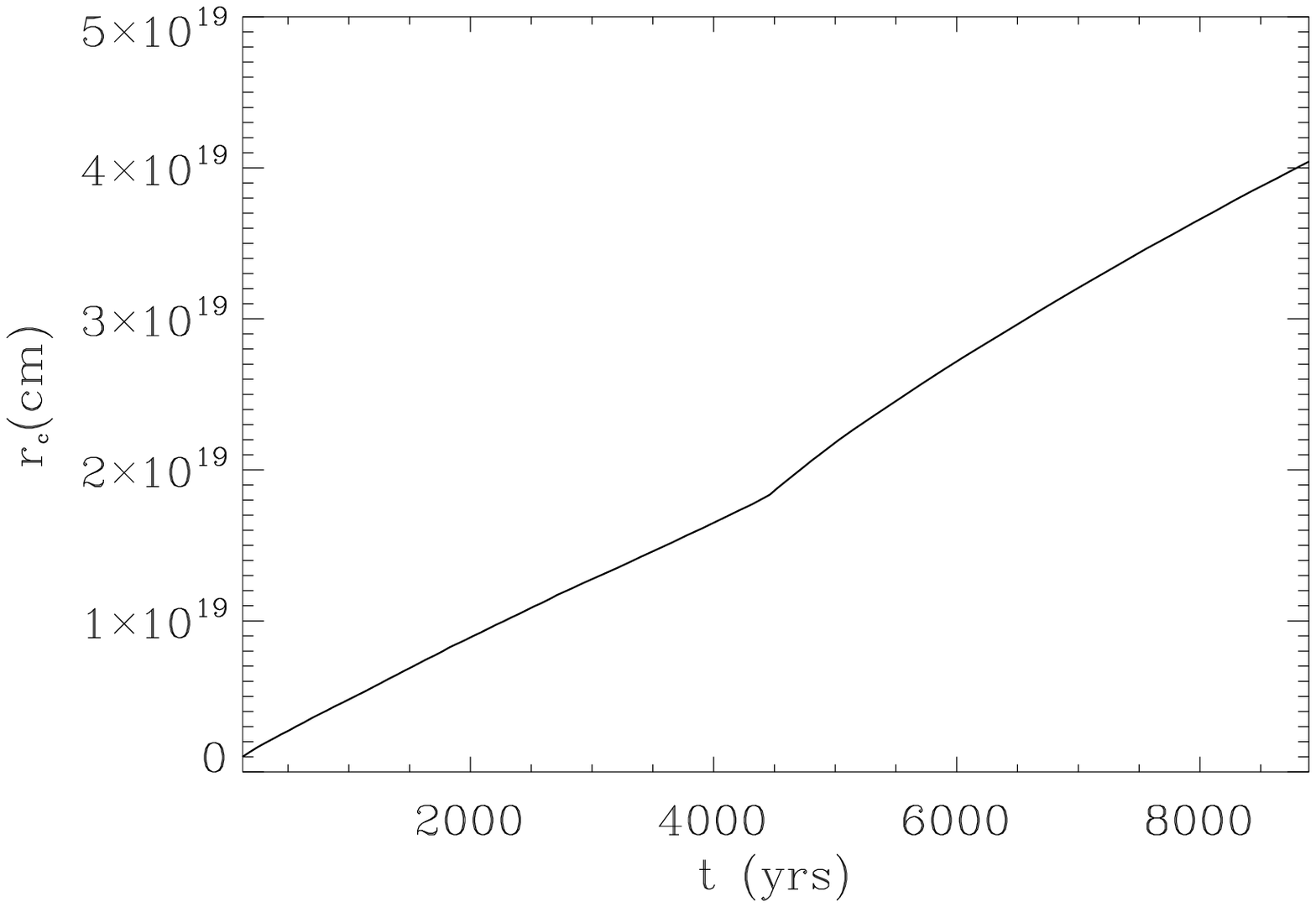}
   \caption{Position of the jet head (left) and mean radius of the bow-shock region (right) for case 1. Note that the bow shock has already left the shown grid by the end of the simultion. Units are cgs.}
   \label{fig:evol1}
   \end{figure}

   \begin{figure}[!h]
     \centering
   \includegraphics[clip,angle=0,width=\columnwidth]{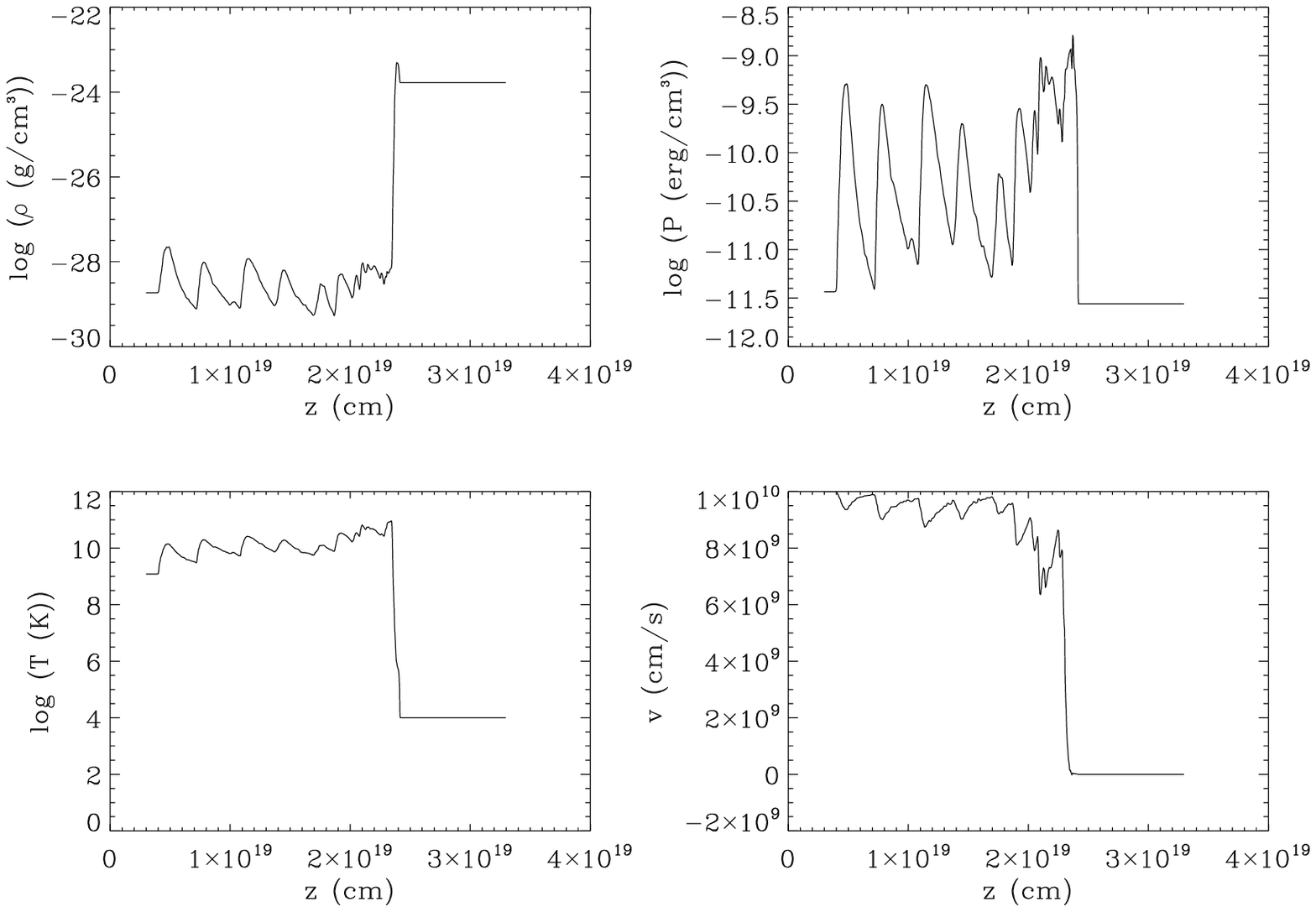}
   \caption{Same as in Fig.~\ref{fig:ax1} but for case 2 ($t_{\rm src}=9800$~yr).}
   \label{fig:ax2}
   \end{figure} 

      \begin{figure}[!h]
     \centering
   \includegraphics[clip,angle=0,width=\columnwidth]{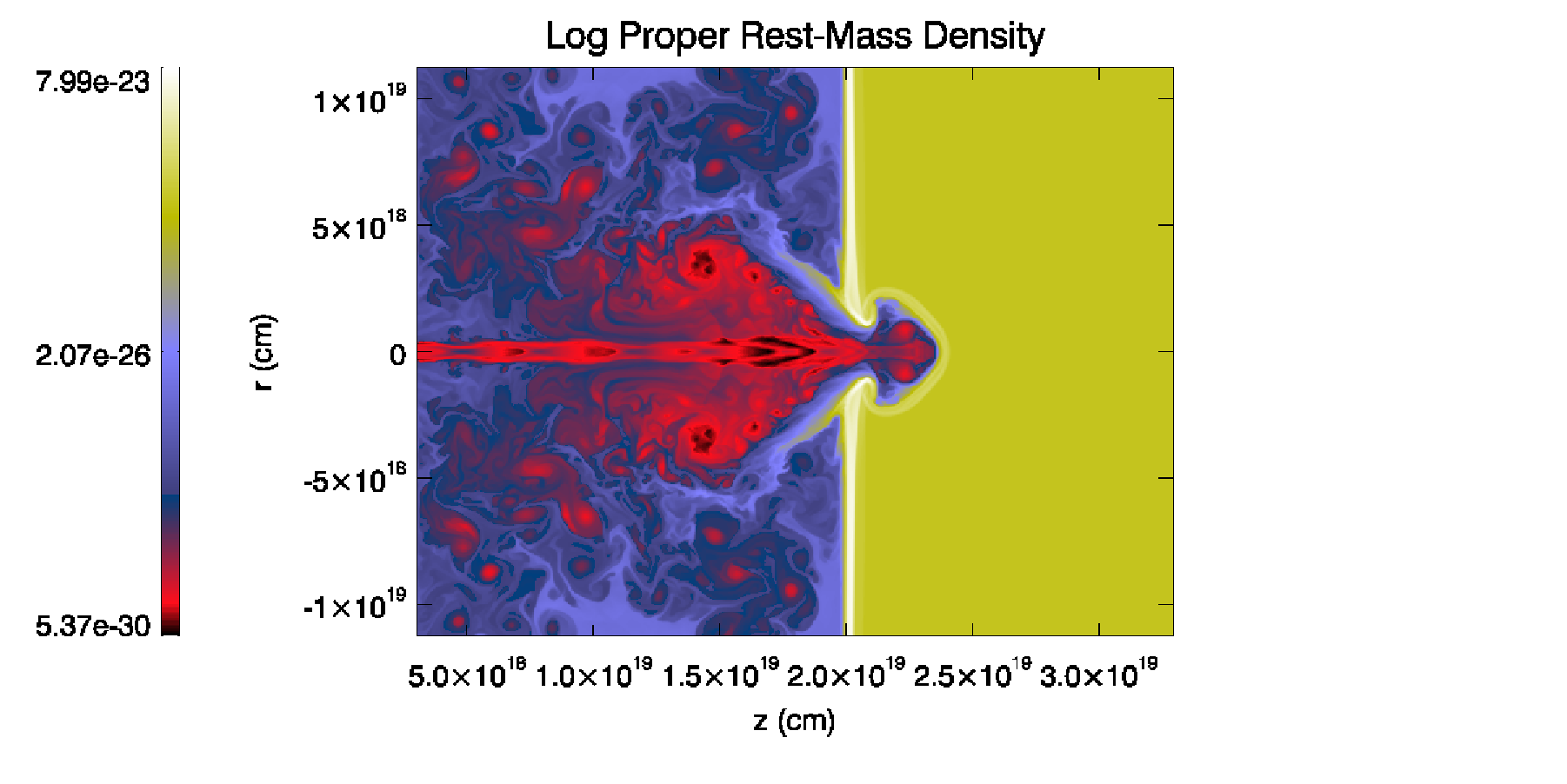}
   \includegraphics[clip,angle=0,width=\columnwidth]{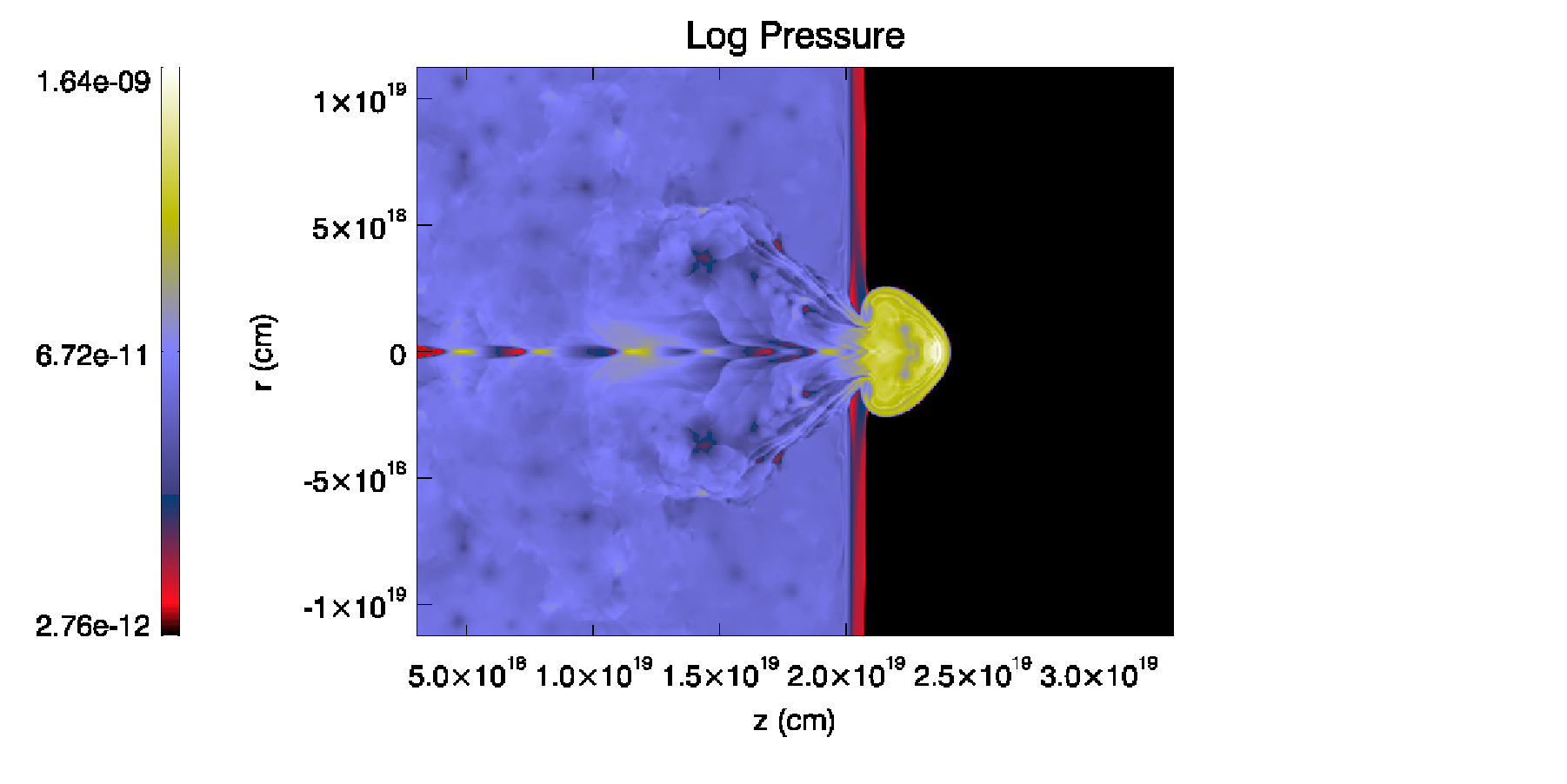}
   \includegraphics[clip,angle=0,width=\columnwidth]{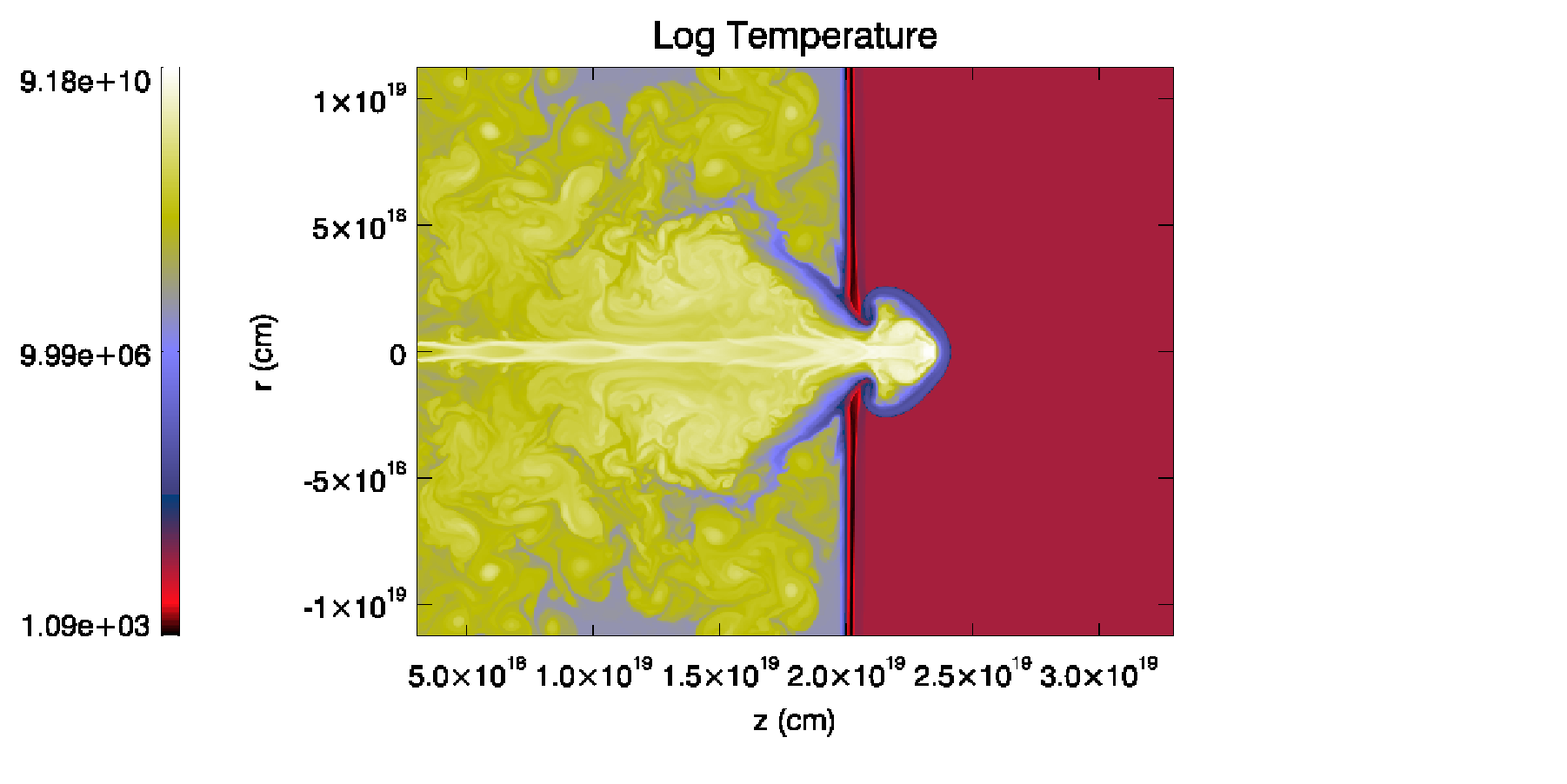}
   \includegraphics[clip,angle=0,width=\columnwidth]{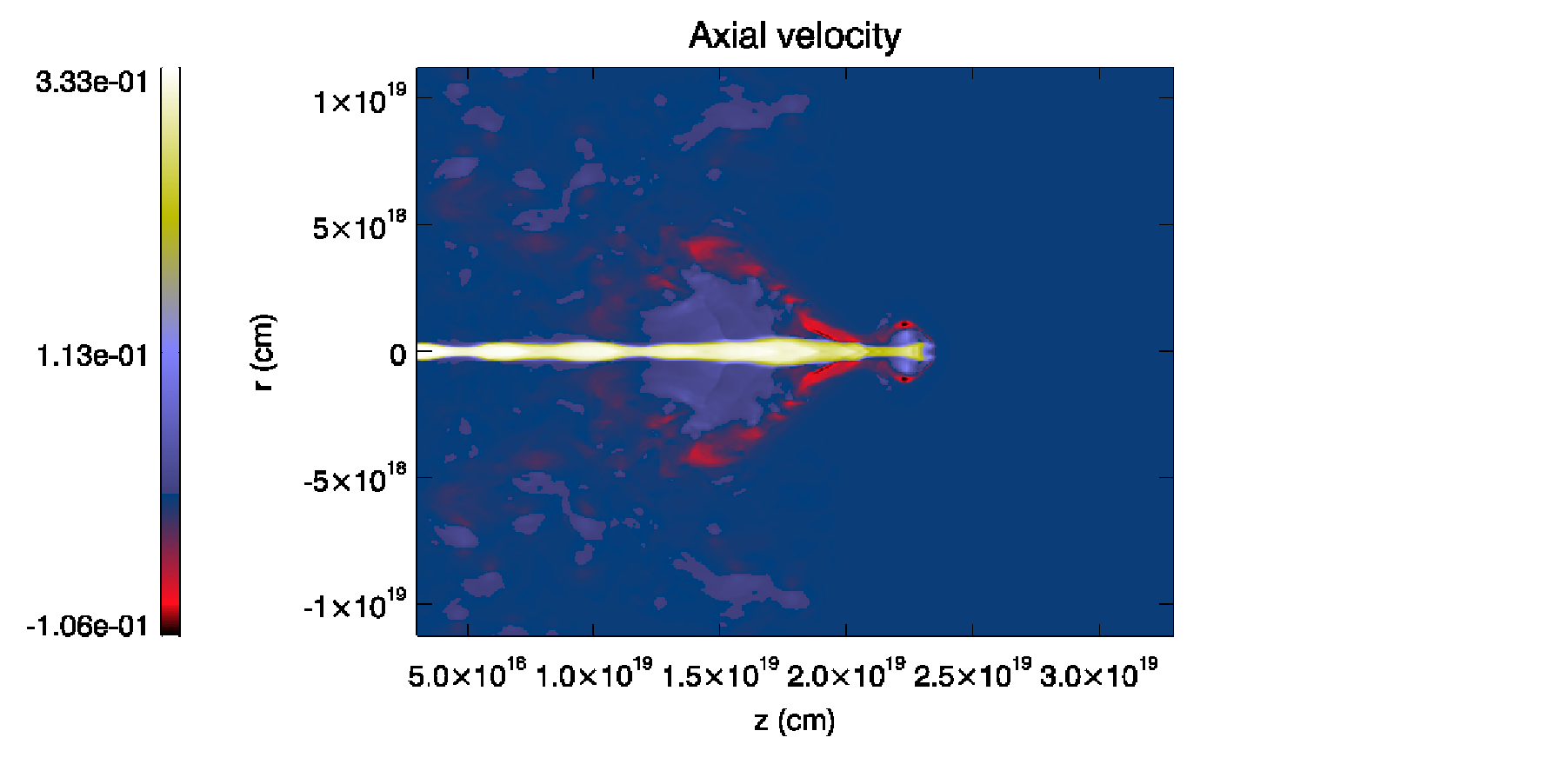}
   \caption{Same as in Fig.~\ref{fig:maps1} but for case 2 ($t_{\rm src}=9800$~yr).}
   \label{fig:maps2}
   \end{figure}

    \begin{figure}[!h]
     \centering
   \includegraphics[clip,angle=0,width=0.45\columnwidth]{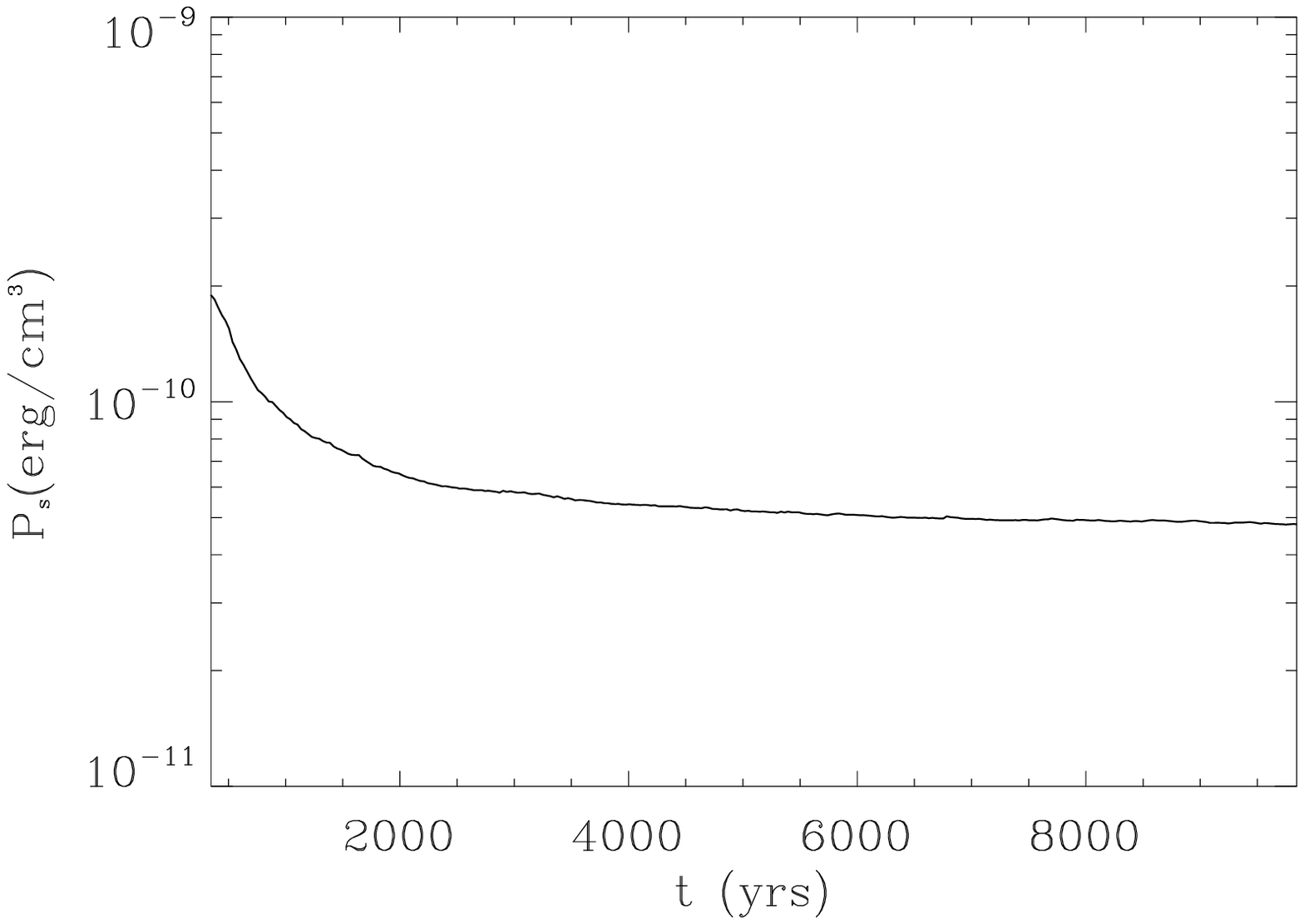}
   \includegraphics[clip,angle=0,width=0.45\columnwidth]{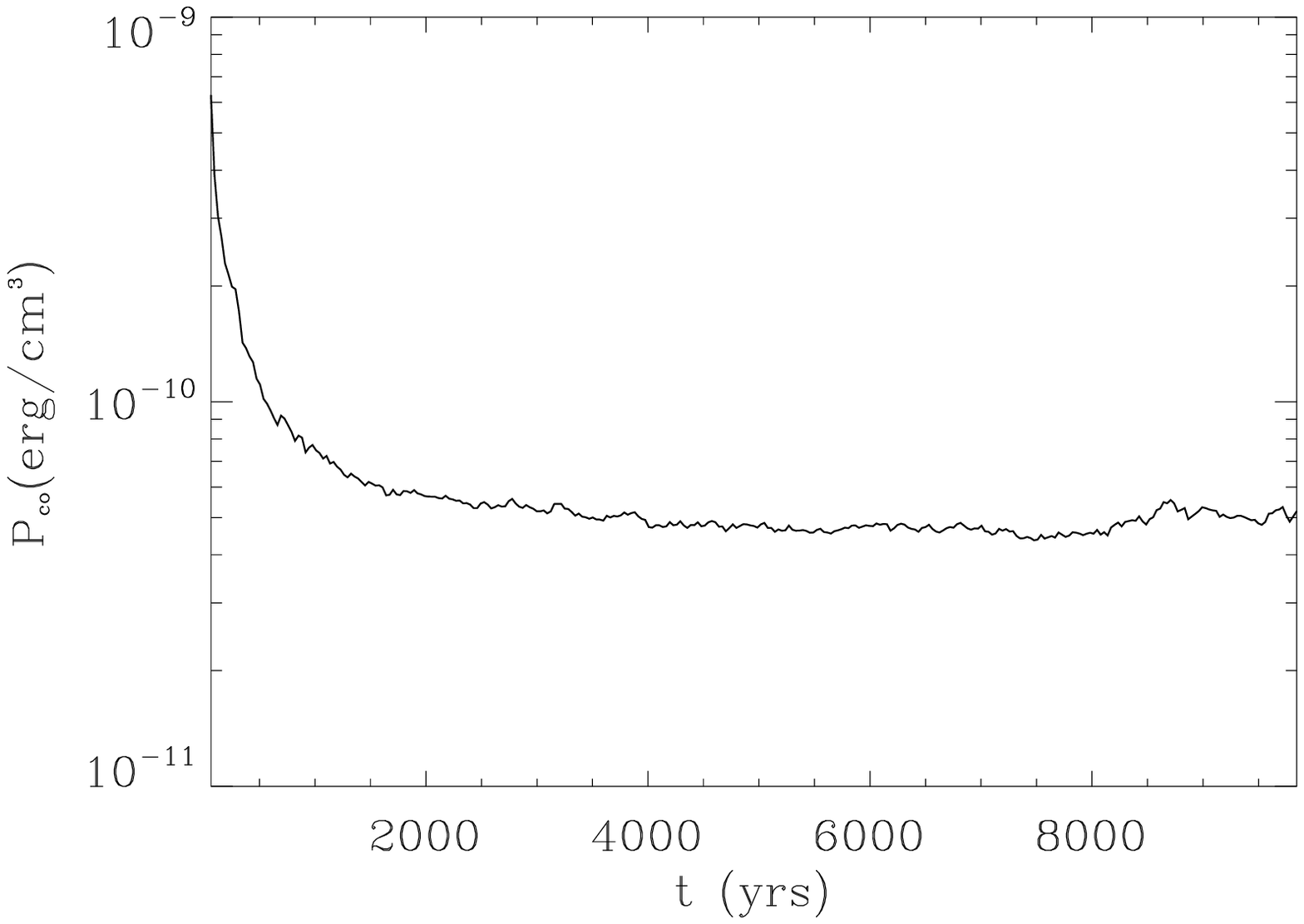}
   \includegraphics[clip,angle=0,width=0.45\columnwidth]{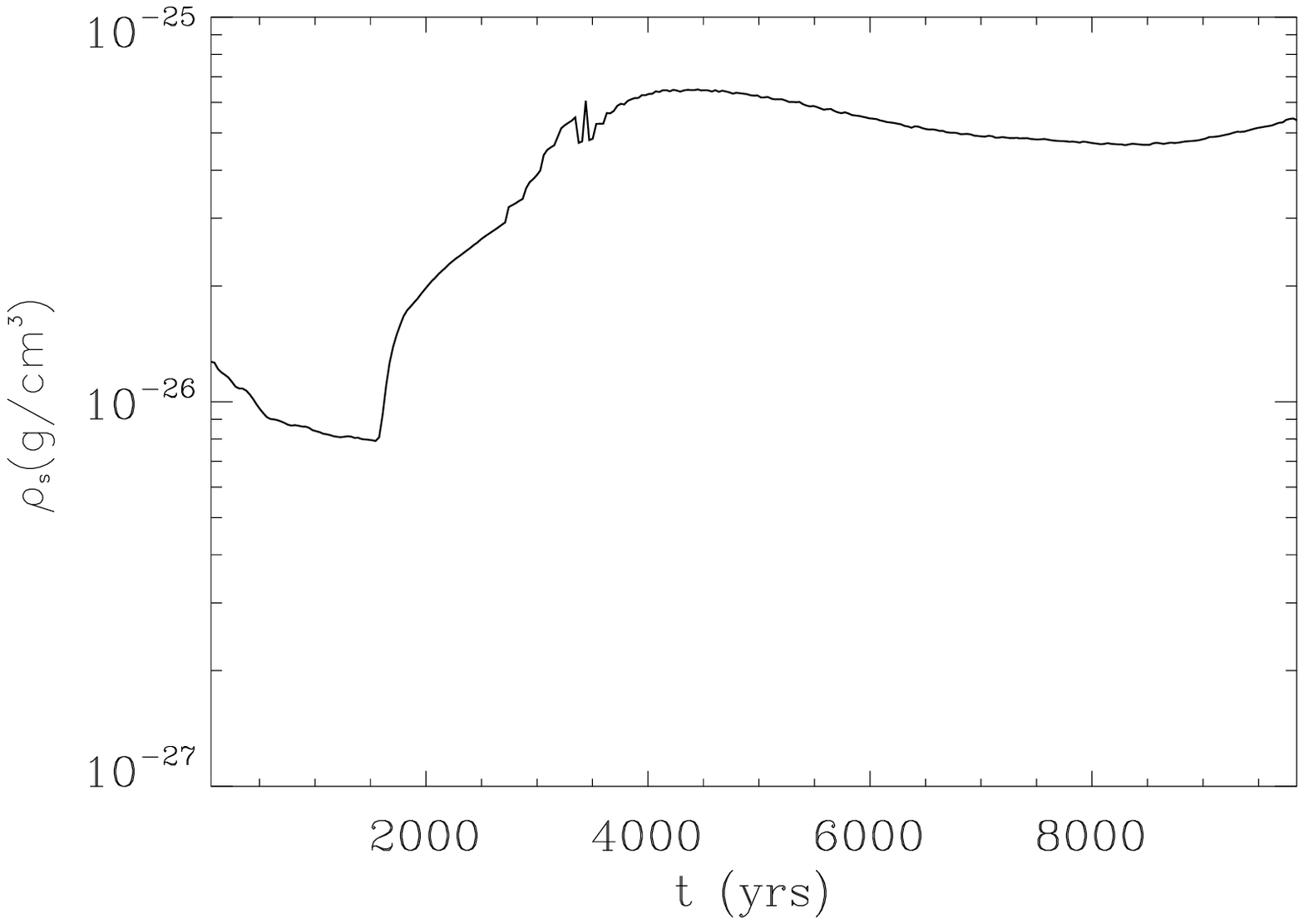}
   \includegraphics[clip,angle=0,width=0.45\columnwidth]{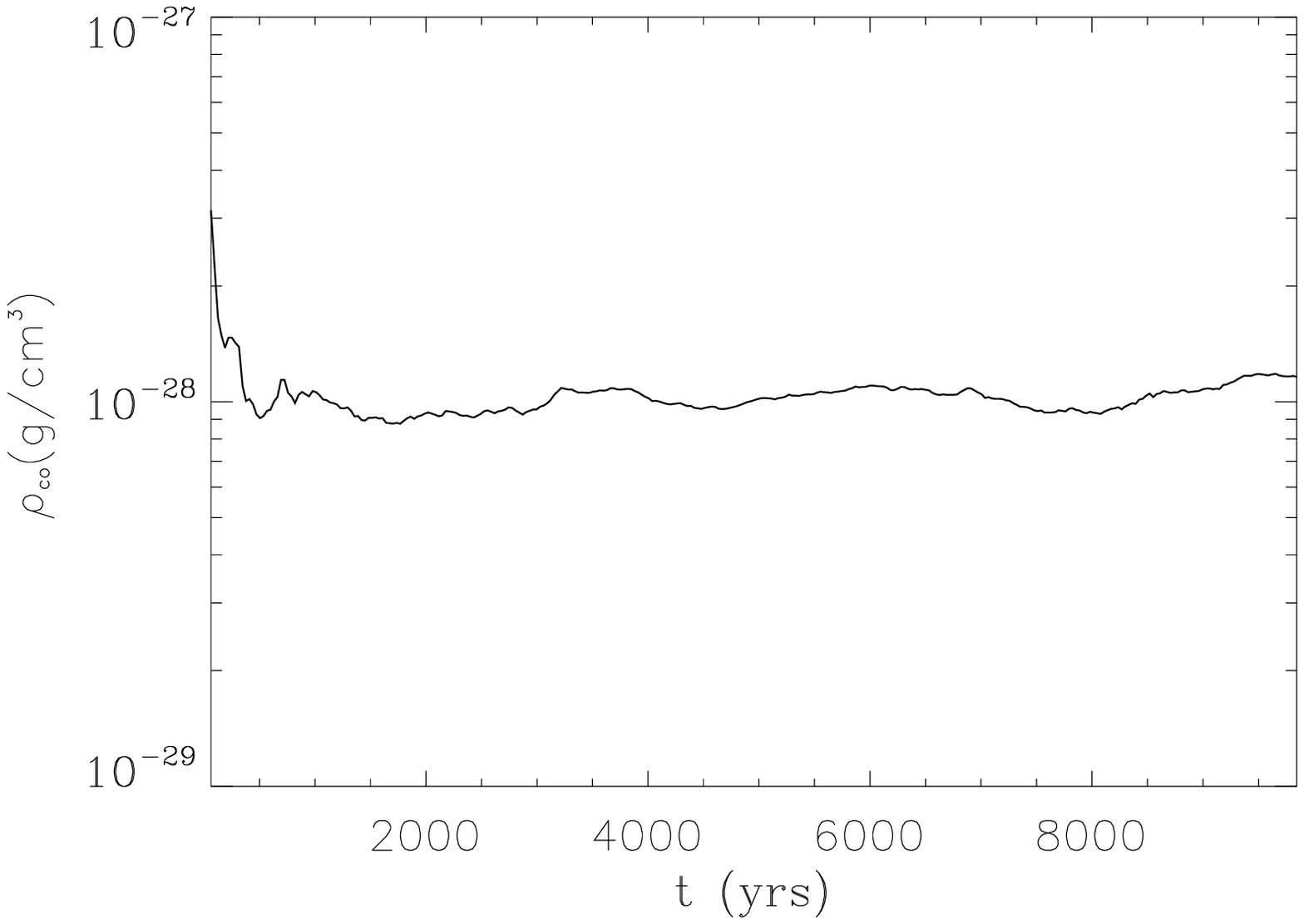}
   \includegraphics[clip,angle=0,width=0.45\columnwidth]{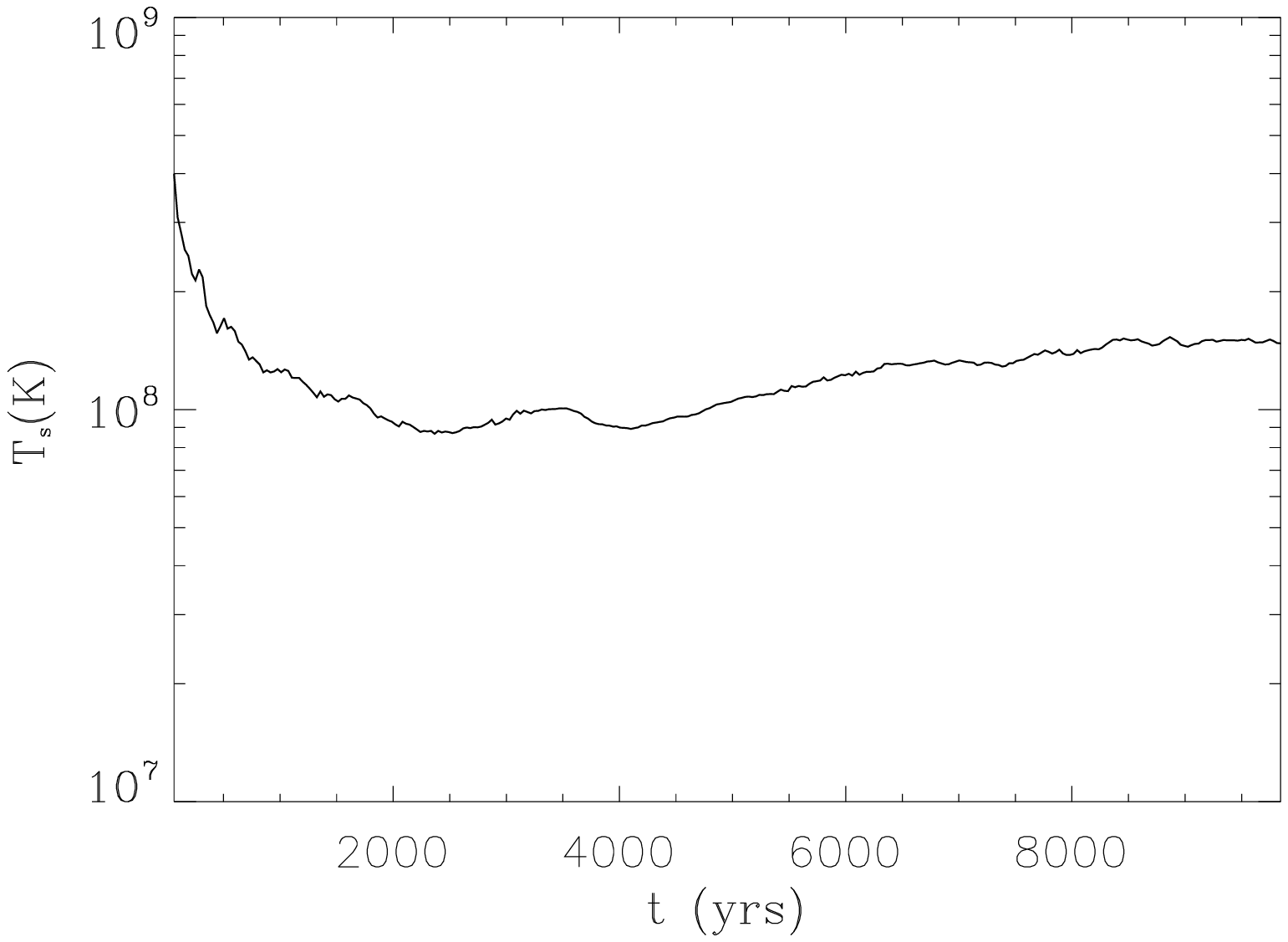}
   \includegraphics[clip,angle=0,width=0.45\columnwidth]{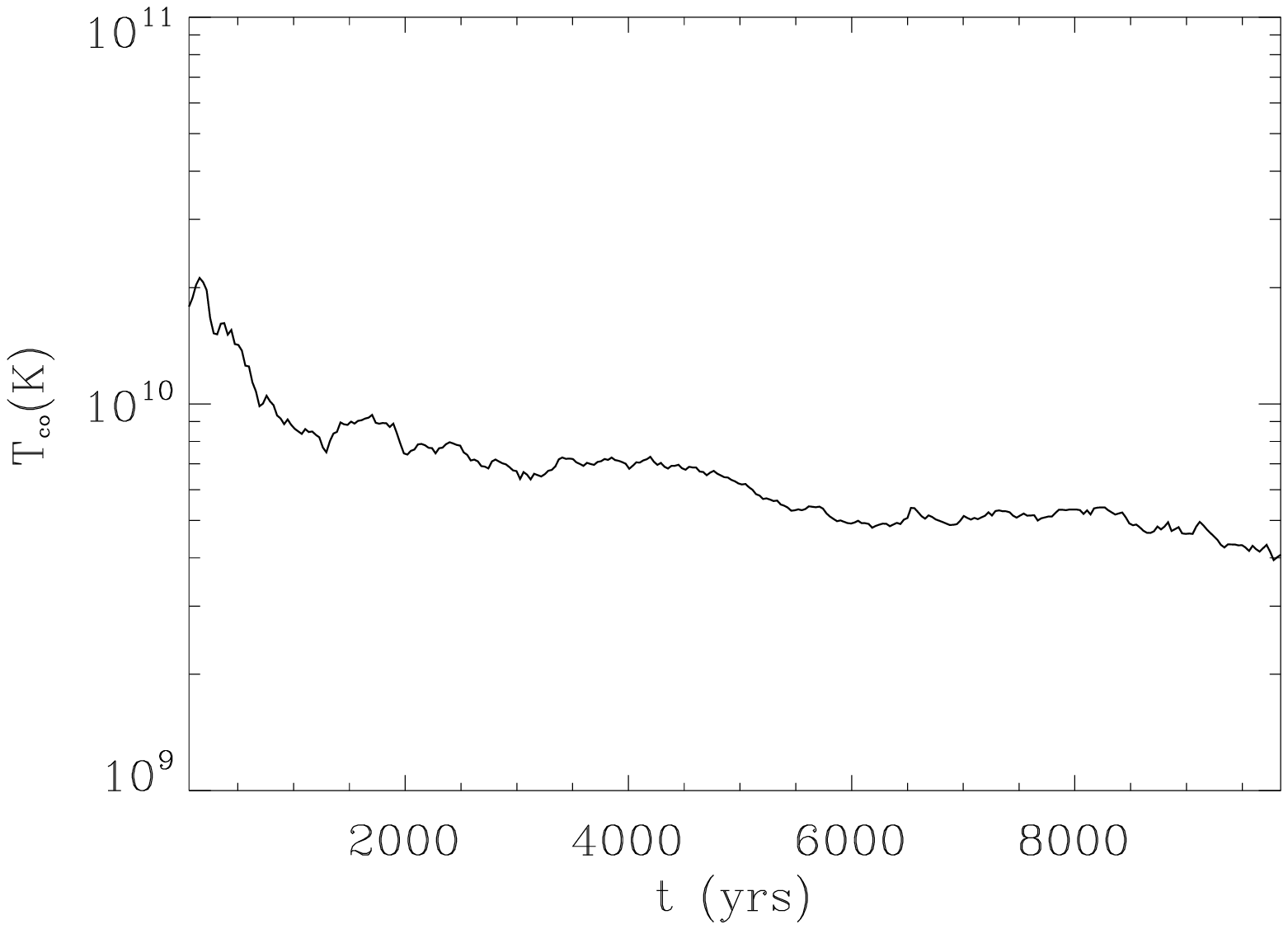}
   \caption{Same as in Fig.~\ref{fig:evol12} but for case 2.}
   \label{fig:evol22}
   \end{figure}

    \begin{figure}[!h]
     \centering
   \includegraphics[clip,angle=0,width=0.45\columnwidth]{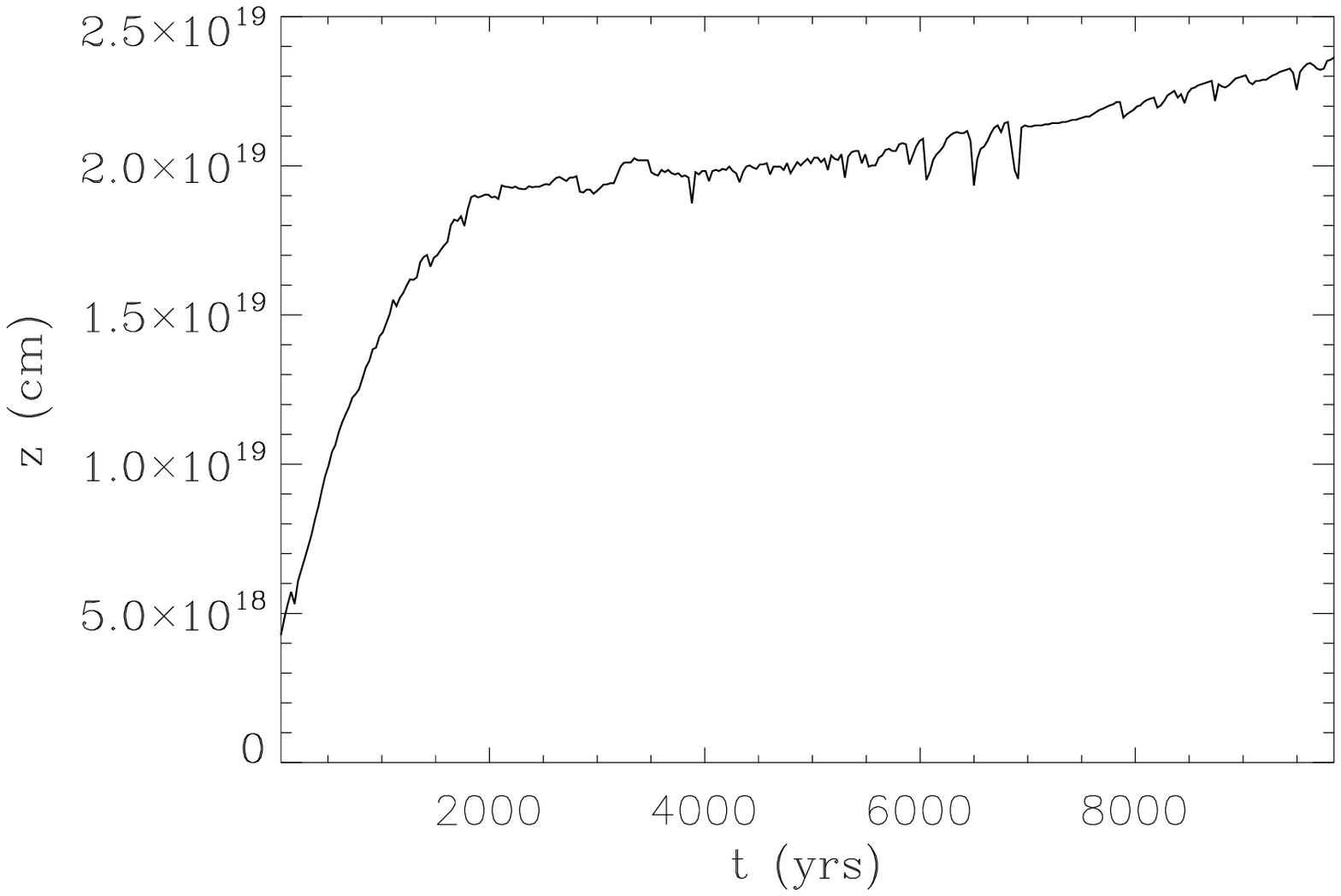}
   \includegraphics[clip,angle=0,width=0.45\columnwidth]{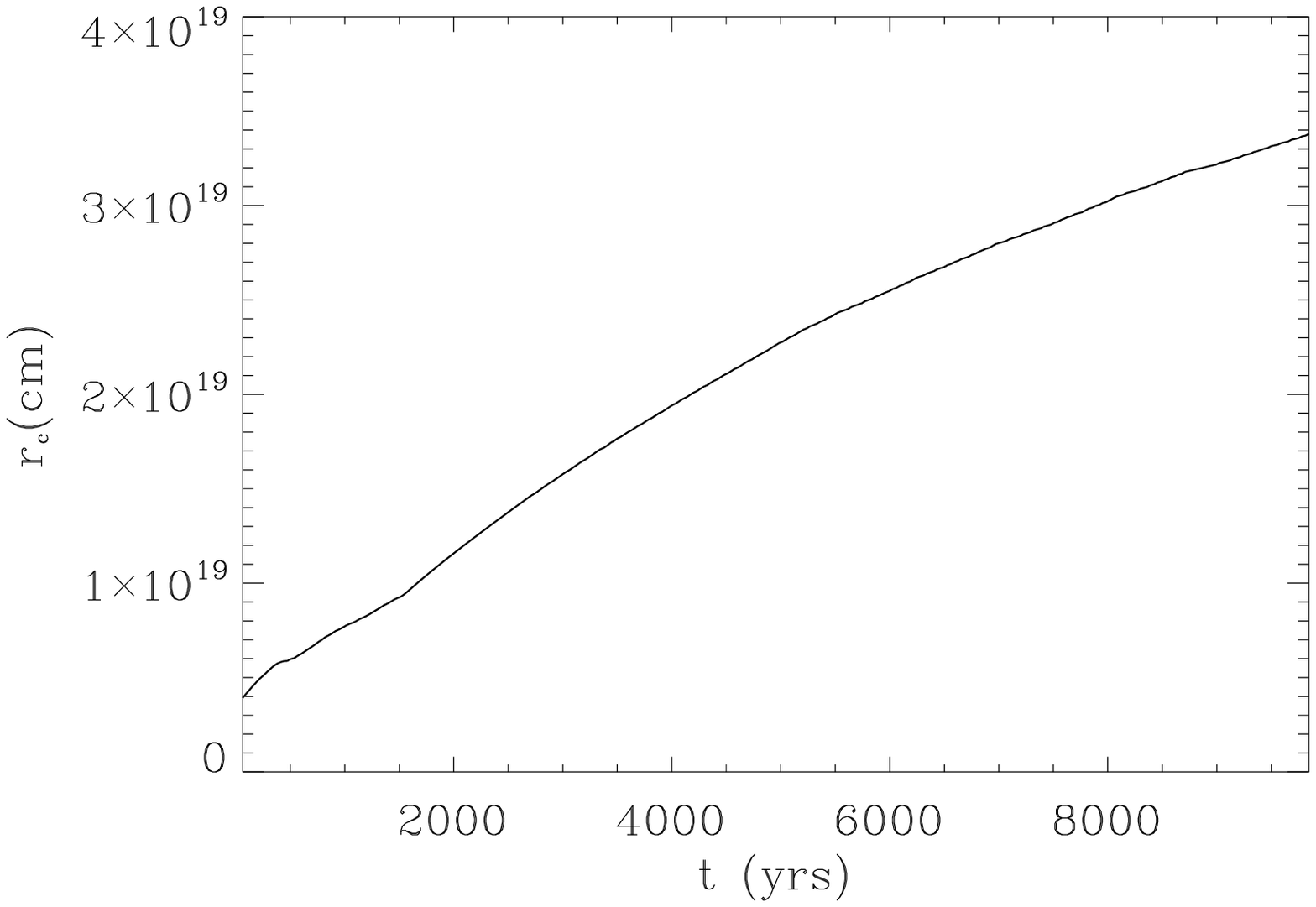}
   \caption{Same as in Fig.~\ref{fig:evol1} but for case 2.}
   \label{fig:evol2}
   \end{figure}

      \begin{figure}[!h]
     \centering
   \includegraphics[clip,angle=0,width=\columnwidth]{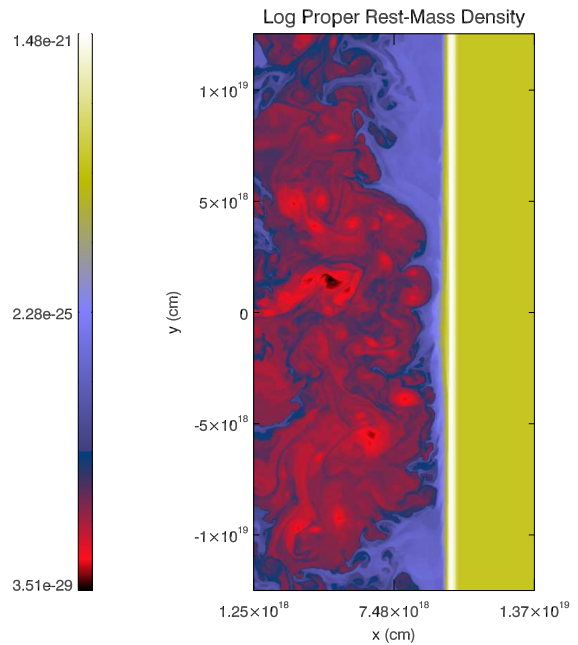}
   \includegraphics[clip,angle=0,width=\columnwidth]{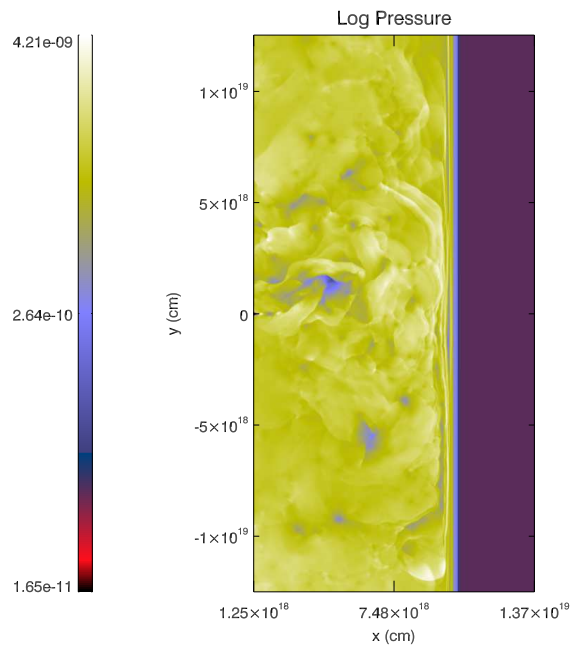}
   \caption{Maps of density (top) and pressure (bottom) at the last snapshot 
   of the evolution for case 3 ($t_{\rm src}\approx 2000$~yr). The shocked wind moves from top to bottom.}
   \label{fig:maps31}
\end{figure}
      \begin{figure}[!h]
     \centering
   \includegraphics[clip,angle=0,width=\columnwidth]{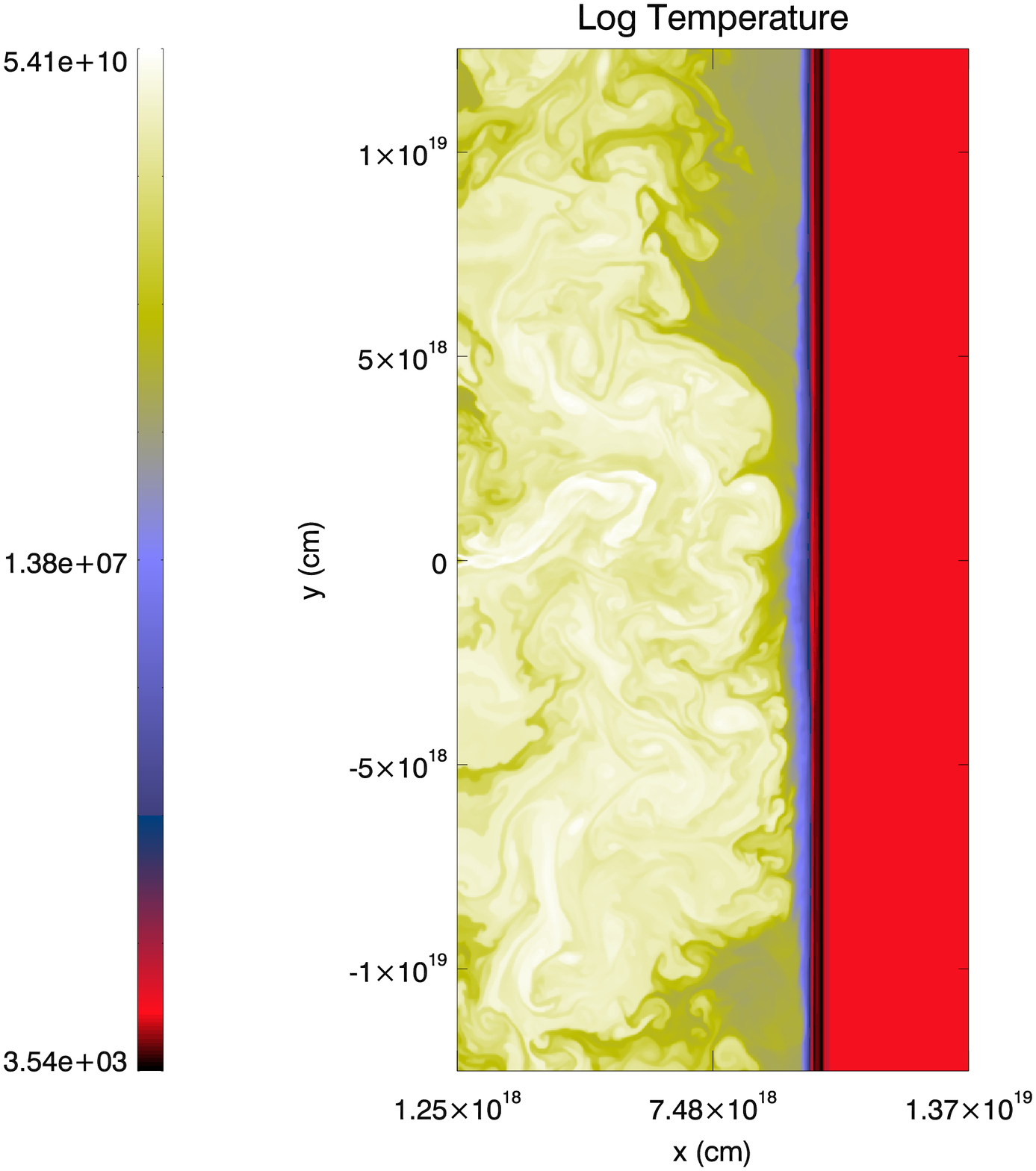}
   \includegraphics[clip,angle=0,width=\columnwidth]{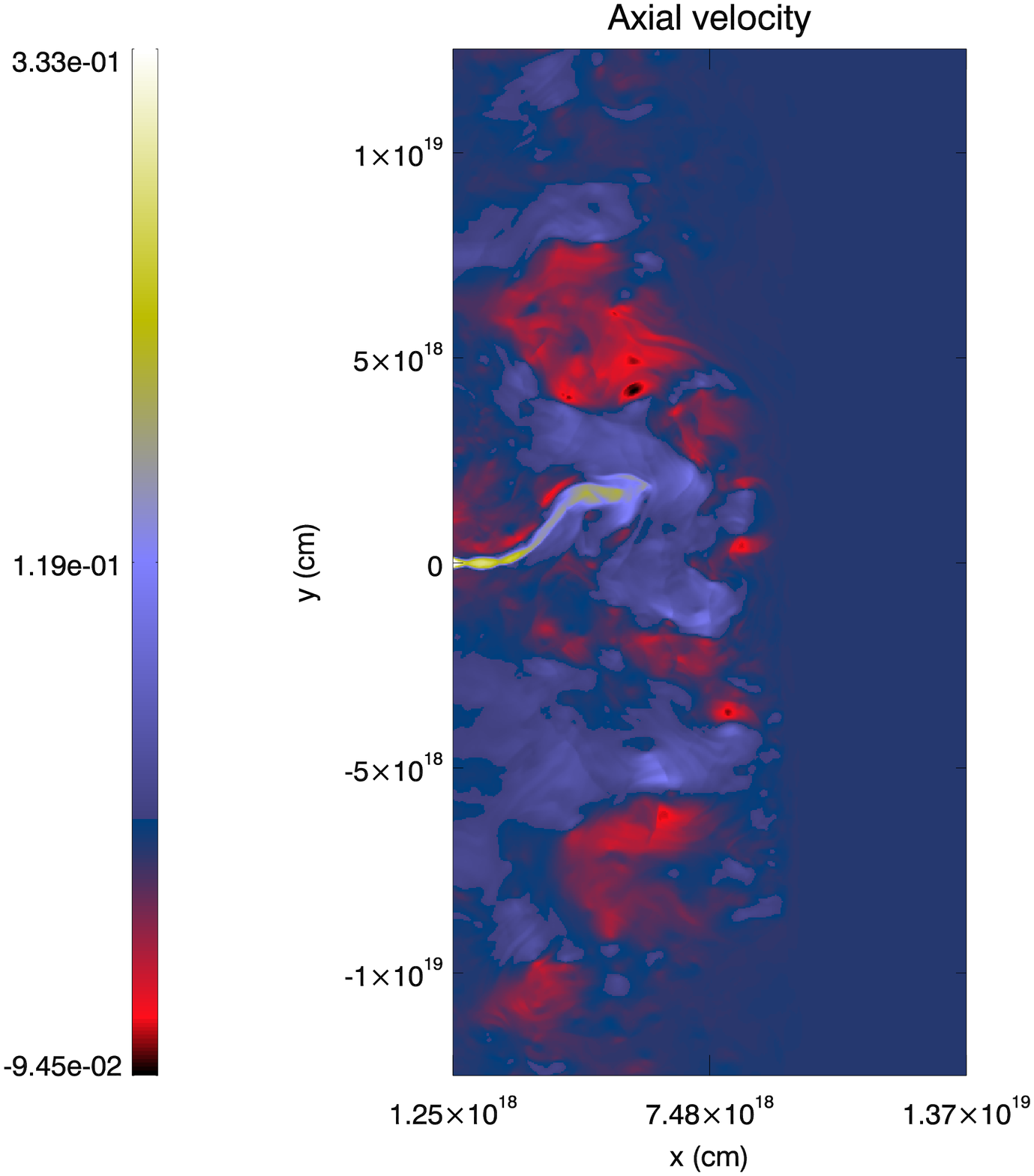}
   \caption{Maps of temperature (top) and axial velocity (bottom) at the last snapshot of the evolution for case 3 ($t_{\rm src}\approx 2000$~yr). The shocked wind moves from top to bottom.}
   \label{fig:maps32}
   \end{figure}

     \begin{figure}[!h]
     \centering
   \includegraphics[clip,angle=0,width=\columnwidth]{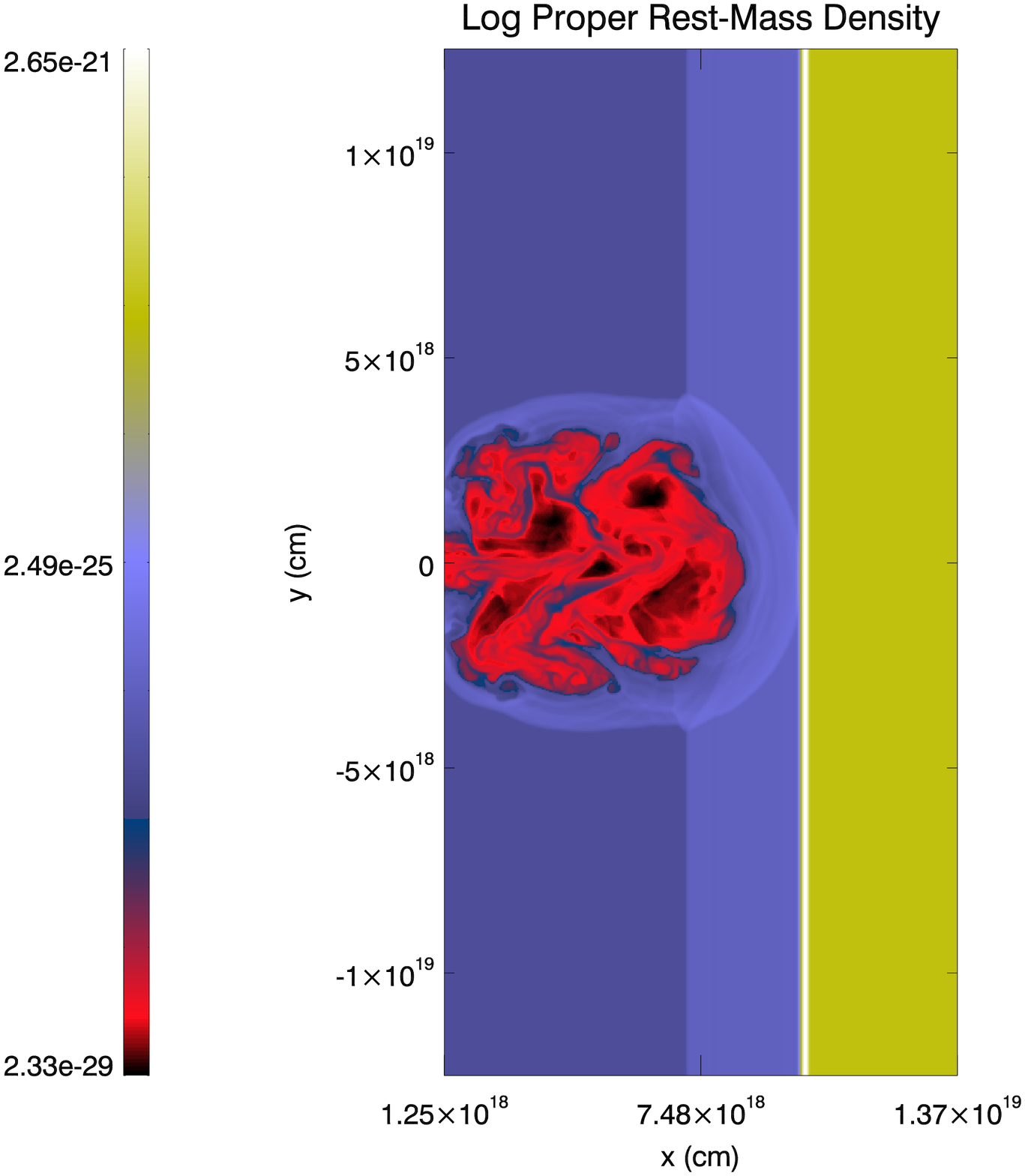}
   \includegraphics[clip,angle=0,width=\columnwidth]{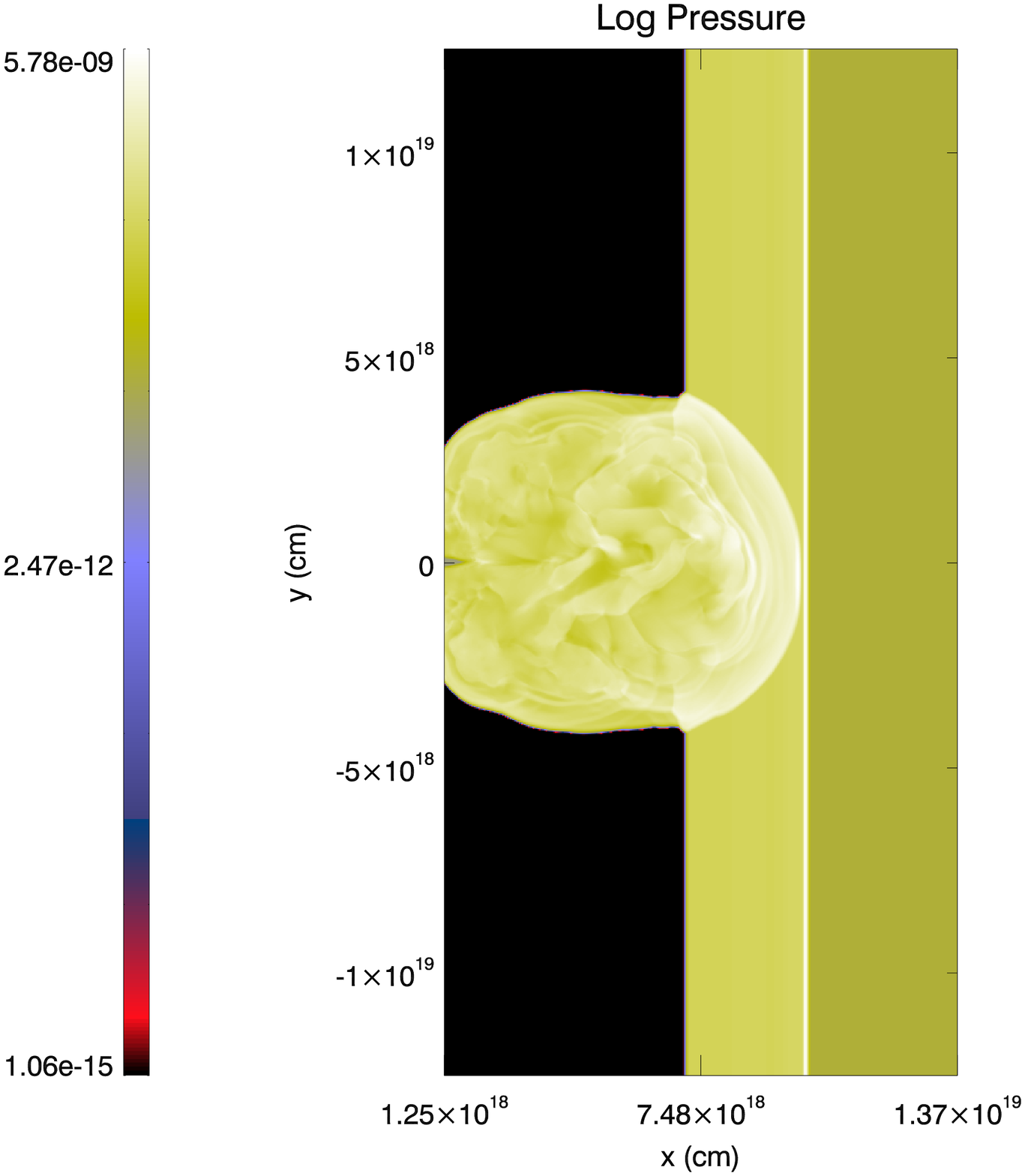}
 \caption{Same as in Fig.~\ref{fig:maps31} but for a source age of 300~yr.}
   \label{fig:maps31b}
\end{figure}
      \begin{figure}[!h]
     \centering
   \includegraphics[clip,angle=0,width=\columnwidth]{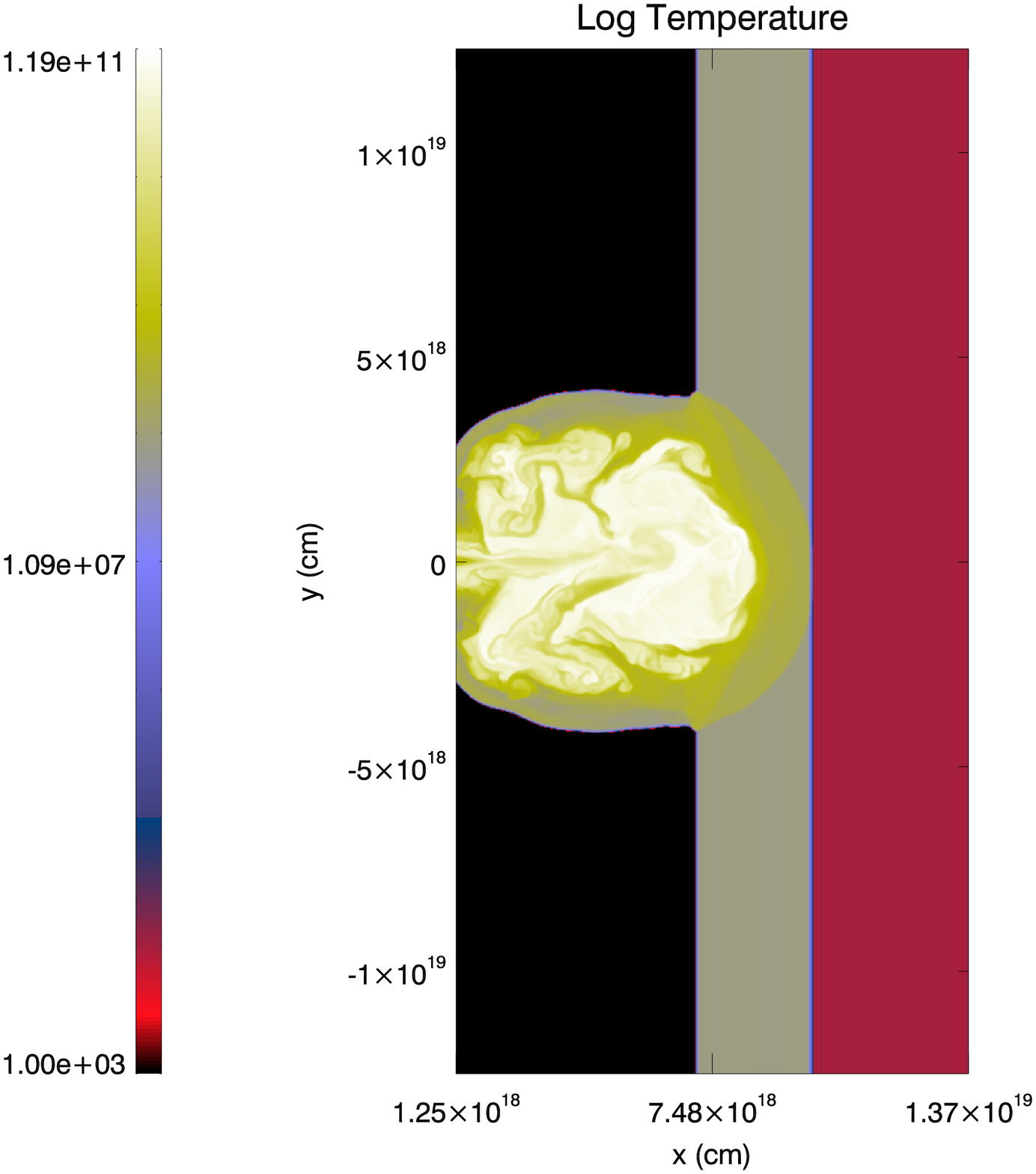}
   \includegraphics[clip,angle=0,width=\columnwidth]{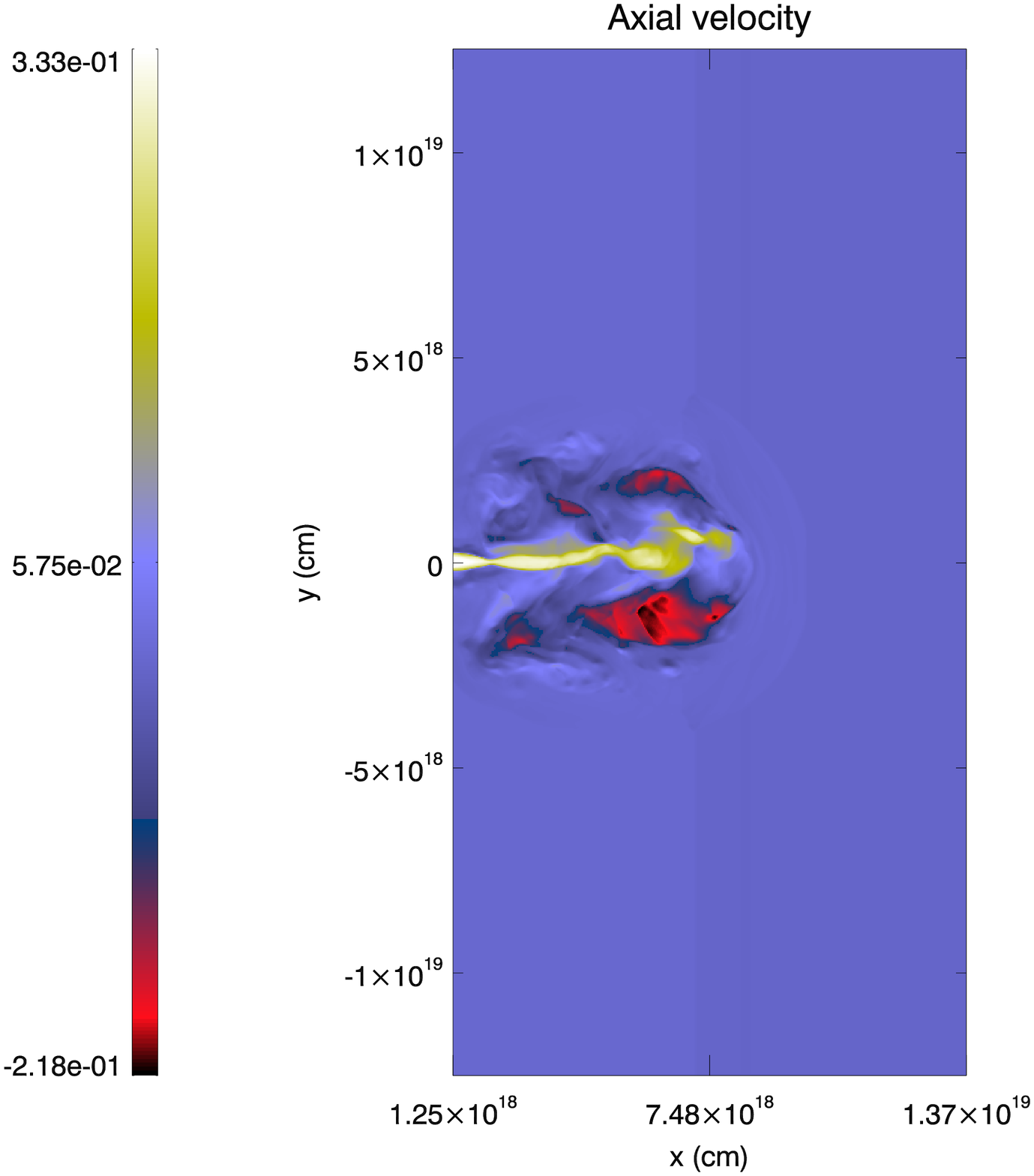}
   \caption{Same as in Fig.~\ref{fig:maps32} but for a source age of 300~yr.}
   \label{fig:maps32b}
   \end{figure}

The simulation of case 1 has been run during 8900~yr of evolution. During the first 4500~yr, the jet propagates up
to $z\approx 3\times 10^{19}\,\rm{cm}$, with an advance velocity $v_{\rm h}\approx 2\times 10^8\,\rm{cm\,s^{-1}}$.
The jet is initially underpressured with respect to the hot ambient medium (shocked wind). This generates a
strong conical recollimation shock in the jet soon after injection, followed by pinching. At $z\approx 3\times
10^{19}\,\rm{cm}$, the jet head reaches the shocked ISM and its advance is stopped. The speed of the head is reduced
to $2\times 10^7\,\rm{cm\,s^{-1}}$ and the jet is only able to propagate up to $z\approx 3.2\times 10^{19}\,\rm{cm}$
by the end of the simulation. The expansion of the bow shock during this phase is mainly radial (perpendicular to
$z$), with an expansion to a distance of  $4.3\times 10^{19}\,\rm{cm}$ from the jet axis (outside the frame of the
image). The reverse shock becomes stronger after the collision with the shocked ISM, and  pinching is enhanced along
the jet. By the end of the simulation, the SNR shock in the ISM has expanded farther, relaxing the density and
pressure conditions that the jet finds at arrival, which makes easier for the jet to carve its way across this
region. At the given velocity, the jet would need about 3000~yr more to complete its path through the
shocked ISM. 

In case 1, the pressure in the cocoon and bow-shock regions is basically constant from the early stages of the
simulation: $P_s \approx 5\times 10^{-10}\,\rm{erg/cm^3}$, whereas the density starts increasing when the jet head
reaches  the ISM shocked by the SNR, from $3\times 10^{-26}\,\rm{g/cm^3}$ to $5\times 10^{-26}\,\rm{g/cm^3}$ during
the period  4500--8900~yr. The temperature  in the shocked ambient (shocked wind and SNR ejecta) falls slowly during
the whole run from $2\times 10^{8}$ to $1.2\times 10^{8}\,\rm{K}$.  In the cocoon, the pressure is very
similar to that in this shocked ambient. The cocoon density increases from  $3\times 10^{-28}\,\rm{g/cm^3}$ to
$7\times 10^{-28}\,\rm{g/cm^3}$ during the first 2000~yr due to the jet low pressure with respect to the ambient, and
remains fairly constant afterwards due to the recollimation and pinching shocks. The temperature in the cocoon
decreases from $3\times 10^{10}$ to $10^{10}$~K during these first 2000~yr, and becomes constant from that
moment on. Turbulence and mixing of jet and external material takes place in the cocoon. Turbulence follows, down to
the scales allowed by the simulation resolution, a spectrum similar to Kolmogorov, 
and the entrained external material actually dominates the mass content of the cocoon.
The axial cuts of density, pressure, temperature and velocity at the last frame, presented in
Fig.~\ref{fig:ax1}, show the rich structure in the jet after the jet is pinched by the ambient pressure, with plenty
of shocks, up to the point where
the jet material reaches the strong reverse shock in  the interaction with the ISM shocked by the
SNR. The recollimation and pinching shocks appear early in the simulation given the low initial lateral pressure of 
the jet with respect to the medium, the pinching structures getting enhanced when the jet head reaches the shocked ISM.
The maps for the density, pressure, temperature and axial velocity at the last snapshot (8900~yr) of the evolution are
shown in Fig.~\ref{fig:maps1}. The temporal evolution of the average cocoon and bow-shock density, pressure,
temperature, and velocity, are shown in Fig.~\ref{fig:evol12}. The temporal evolution of the jet head position and the
mean radius of the bow shock are shown in Fig.~\ref{fig:evol1}.

The simulation of case~2 has been run during 9800~yr of evolution. The jet crosses the stellar wind and shocked wind region,
from injection at $z\approx 3\times 10^{18}\,\rm{cm}$ to the contact discontinuity with the shocked ISM at $z\approx 2\times
10^{19}\,\rm{cm}$, at 1500~yr after injection, at a mean velocity of $v_{\rm h}\approx 3.5\times 10^8\,\rm{cm\,s^{-1}}$. As in
case 1, the jet is initially underpressured and collimated by the ambient, which produces pinching oscillations along the jet
body. After reaching the shocked ISM, the jet head gets flattened due to the large density wall. At that point, the pressure in
the bow-shock and cocoon regions of the jet starts to decrease slower with time. The same happens with the density in the
cocoon,  whereas the density of the shocked ambient starts to increase, since the head of the jet interacts with a denser
medium. The backflow expands sideways, filling the simulated region, whereas the head of the jet acts as a drill on the shocked
ISM region due to its much higher pressure and crosses it at $t\approx 4300$~yr after injection. Thus, it needs 3000~yr to
cross this small shell. As in case 1, the internal structure of the jet becomes very rich due to the interaction between the
previous pinching patterns and the backward moving waves produced at the shock of the jet flow with the stagnation point, at
$z\approx 2.4\times 10^{19}\,\rm{cm}$. The axial cuts of the jet, presented in Fig.~\ref{fig:ax2}, show this effect and the
increase of the temperature and decrease of the velocity after each collimation shock, in which strong pressure jumps are
produced. 
After the jet head crosses the shocked ISM layer, it propagates in the ISM with a supersonic speed of $\approx 2.4\times
10^7\,\rm{cm\,s^{-1}}$, an order of magnitude smaller than that in the shocked wind region. 
The maps for the density, pressure,
temperature and axial velocity at the last snapshot (9800~yr) of the evolution are shown in Fig.~\ref{fig:maps2}. The
temporal evolution of the average cocoon and bow-shock density, pressure, temperature, and velocity, are shown in
Fig.~\ref{fig:evol22}. The temporal evolution of the jet head position and the mean radius of the bow shock are shown in
Fig.~\ref{fig:evol2}.
Overall, the jet evolution of case~2 is similar to that of case~1, 
with the cocoon also affected by turbulence and strong mixing.

Simulation 3 has been run during $\approx$ 2000~yr. The jet is injected in the grid in the wind region, at
$z=1.25\times 10^{18}\,\rm{cm}$. When the jet reaches the shocked wind region, at $z=8\times
10^{18}\,\rm{cm}$, it gets the impact from the shocked wind that has been deflected by the propagation of
the binary system in the ISM (see Fig.~\ref{casos}, bottom). This wind is enough, in the slab simulation,
to distort the bow shock of the jet and disrupt the latter as follows. After crossing the wind/shocked
wind  discontinuity, the bow shock is deviated to the right from injection in the images due to the shocked
wind coming from top to bottom. The backflow is also deviated in this direction due to the generated
pressure gradient. This generates an excess of backflow material in the unshocked wind region at the right
of the jet, resulting in a deviation of the jet to the left. The combination of this process with the
shocks generated at the different discontinuities destroy the jet even inside the wind region.  When the
head of the jet reaches after 300~yr the shocked ISM region, at $z=10^{19}\,\rm{cm}$, the jet has not
enough thrust to penetrate this layer and all the injected flow is thus converted into a backflow that
fills the inner regions and also the unshocked wind region. The combination of disruption and advection
does not allow the jet to reach the normal ISM regardless the time elapsed, and thus jets with the given
power will be trapped inside the wind/shocked wind region, and advected backwards with respect to the
system motion. The maps for the density, pressure, temperature, and axial velocity, at the last snapshot (2000~yr)
and after 300~yr of evolution, are shown in Figs.~\ref{fig:maps31}, \ref{fig:maps32}, \ref{fig:maps31b} and
\ref{fig:maps32b}.

\section{Non-thermal activity}\label{nt}

In order to estimate what may be the potential non-thermal luminosity of the radiation produced when the jet head propagates
within the SNR or the shocked stellar wind, we have used a model for the cocoon emission similar to that  presented in  Bordas
et al. (2009). In short, the model assumes the cocoon as an homogeneous emitter, which is true to first order provided that the
flow is roughly sonic. Our simulations show turbulent motion and jet/medium mixing in the cocoon, and the flow becomes
trans-sonic far from the jet tip, but at this stage we will neglect this effect. 
In the radiation calculations, following
Bordas et al. (2009), the magnetic field ($B$) energy density ($u_{\rm B}$) has been assumed to be a 10\% of the matter energy
density, and the non-thermal injection luminosity, a 1\% of the jet power. Postshock magnetic field values close to, or
not far below, equipartition are expected if the jet keeps a non-negligible fraction of its total energy as magnetic energy,
which is a reasonable assumption provided that jets are magnetically launched (e.g. Komissarov et al. 2007; Spruit 2010).
Note that for an optically thin emitter, the radiation luminosity will be just proportional to this fraction. Unlike in Bordas
et al. (2009), here only the cocoon has been considered as a powerful emitter, since the recollimation and reverse shocks
suffered by the jet are strong, but the bow shock is too weak to accelerate particles. The results for the cocoon radiation are
presented in Sect.~\ref{coc}, in which we focus on electrons only given the low cocoon densities that prevent efficient
gamma-ray production from proton-proton interactions. These low densities imply also that relativistic Bremsstrahlung will be
inefficient there. 

When the jet head impacts the shocked ISM thin shell, the highest energy particles accelerated in the cocoon region may
reach this shell, which in case 2 reaches densities $n_{\rm ISMs}\sim 100$~cm$^{-3}$. Given the relativistic Bremsstrahlung
(electrons) and proton-proton interaction (if protons are also accelerated) timescales and the source age, about a 1--10\%
of the luminosity in relativistic particles reaching that shell could be radiated at very high energies. We note that
denser media would increase the efficiency, which is simply $\propto n_{\rm ISMs}$. 

For the case when the jet has reached the undisturbed ISM, if it is not disrupted, the non-thermal behavior can be
described as in Bordas et al. (2009). This would apply to the emission produced during the second half of case 2
simulation. We have not taken into account in this work particle acceleration in the turbulent and sheared flow in
the recollimated jet and cocoon regions, but stochastic and shear acceleration may be relevant there (see Rieger et
al. 2007 and references therein). Some thermal radiation may be produced in the shocked ISM. This thermal component
tends to be quite extended and, given the shock velocities, would peak in the UV.

We do not consider here case 3 since, due to jet full disruption, conditions seem less suitable for particle
acceleration. It might be that stochastic and shear acceleration may still take place in the jet disrupted medium,
although estimating the non-thermal emission for this case requires an in-depth study that is beyond the scope of
this work.

\subsection{Cocoon emission: cases 1 and 2}\label{coc}

\begin{figure*}[]
\centering
\includegraphics[width=0.9\textwidth]{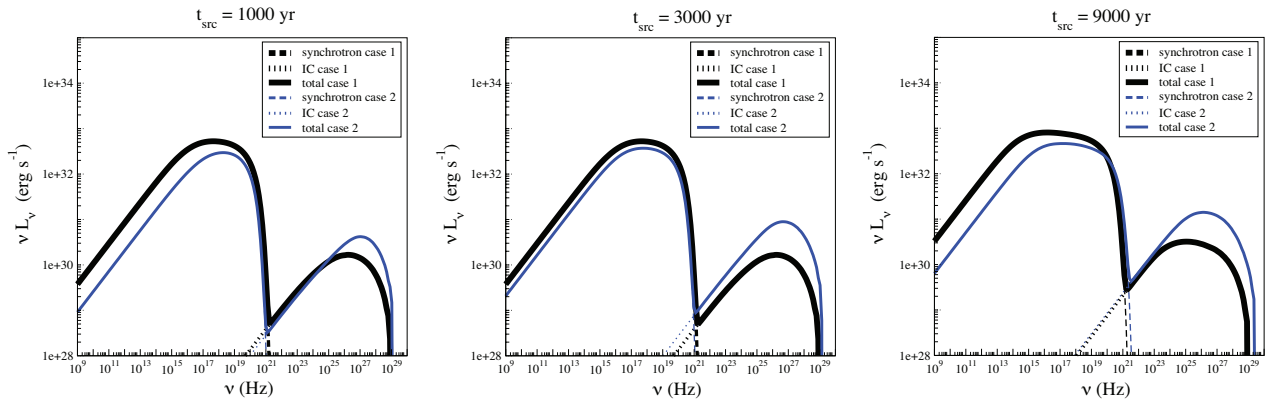}
\caption{Synchrotron and IC SEDs for both simulated cases 1 (black/thick lines) 
and 2 (blue/thin lines) computed for $t_{\rm src}=10^{3}$ (left), $t_{\rm src}=3\times 10^{3}$ (middle) and $t_{\rm src}=9\times 10^{3}$~yr (right).}
\label{fig:SEDs}
\end{figure*}

\begin{table}
\begin{center}
\caption{Model parameters for the shell and cocoon for three different source ages.}
\label{tab:model}
\begin{tabular}{@{}lc}
\hline
\textit{Parameter}   &  value \\
\hline
Jet power [erg~s$^{-1}$] &  $3 \times 10^{36}$   \\
Non-thermal power fraction & 0.01 \\
stellar field energy density (\textit{case 1}) [erg~cm$^{-3}$] & $3.1\times 10^{-12}$ \\
stellar field energy density (\textit{case 2}) [erg~cm$^{-3}$] & $5.2\times 10^{-12}$ \\
magnetic field (\textit{case 1}) [G] & $4.1\times 10^{-5}$ \\
magnetic field (\textit{case 2}) [G] & $1.4\times 10^{-5}$ \\
Acceleration efficiency & 0.02 \\
Injection power-law index & 2.1 \\
Maximum energy (\textit{case 1}) & $10^{3}$~TeV\\
Maximum energy (\textit{case 2}) & $1.8 \times 10^{3}$~TeV\\
\hline
\end{tabular}
\medskip
\end{center}
\end{table}

The simulation results are used to characterize the physical properties of the shocked regions. The position and
velocity of the bow shock, the internal energy density ($u_{\rm th}$; and hence $u_{\rm B}=0.1\,u_{\rm th}$), the
temperature, and the rest mass density are used at any source age to account for the evolution of the particle
energy distribution, which is affected by both adiabatic and radiative losses: relativistic Bremsstrahlung
(negligible), inverse Compton (minor) and synchrotron (dominant). For the IC emission, the radiation field provided
by the companion star dominates over that of the CMB. The star is taken to be a bright O, with luminosity
$10^{39}$~erg~s$^{-1}$ and temperature $4\times 10^4$~K. At a distance of $\sim 10$~pc, the energy density ratio is
$u_*/u_{\rm CMB}\sim 7$, increasing towards the central engine like $\propto z^{-2}$. 

We have assumed that the recollimation and reverse shocks accelerate particles, simplifying them as just one accelerator with
{\it average} properties. The shock velocity has been taken $v_{\rm s}\sim v_{\rm j}$, which yields an acceleration efficiency
$\dot{E} \sim (1/2\pi)\,(v_{\rm s}/c)^{2}\,q\,B\,c\approx 0.02\times q\,B\,c$ (e.g. Drury 1983). As mentioned before, the
assumed non-thermal injection luminosity is a 1\% that of the jet, i.e. $3\times 10^{34}$~erg~s$^{-1}$, and the magnetic to
internal energy density ratio is 0.1. This gives a $B$-strength of about $\sim 4-6\times 10^{-5}$~G depending on $t_{\rm src}$.
These $B$-values are one order of magnitude higher than those inferred by Tudose et al. (2006) and Safi-Harb \& Petre
(1999) for the jet termination regions in Circinus~X-1 (a low-mass microquasar) and SS~433, respectively. On the other hand, the ages of these
sources are also about one order of magnitude older than those considered here, and thus a lower magnetic field is naturally expected in
the former cases. Under the adopted magnetic field, synchrotron losses are dominant in the cases studied here. For the
conditions assumed, maximum energies are typically $\sim 10^{3}$~TeV. We remark that the adopted non-thermal fraction is quite
conservative, and there is room to increase it. The values of the radiation model parameters are listed in
Table~\ref{tab:model} for cases 1 and 2 when the source age is $\sim 10^4$~yr.

The spectral energy distributions (SED) of the non thermal synchrotron and IC emission for cases 1 and 2, for evolution
times $t_{\rm src}= 10^{3}, 3 \times 10^{3}$ and $9 \times 10^3$~yr, are shown in Fig.~\ref{fig:SEDs}. Depending on the age
of the source and the case, and given that $u_{\rm B}\gg u_{\rm *}$, the synchrotron luminosity can be between $\sim
10-1000$ times higher than the IC one. This difference is more pronounced in case 1. This is a result of the smaller amount
of energy pumped by the stellar wind compared to that of a SNR, which leads to a lower pressure and therefore a lower $B$
in case 2. We note that the IC dominance may be only possible if the $B$-strength were well below equipartition. At the end
of the simulations ($t_{\rm src}=9\times 10^{3}$~yr), the synchrotron fluxes are, for a 3~kpc distance source, of $\sim
20$~mJy (case 2) and 100~mJy (case 1), and $\sim 4$ (case 2) and $7\times 10^{-12}$~erg~cm$^{-2}$~s$^{-1}$ (case 1), in
radio and X-rays, respectively. The IC emission yields fluxes of $\sim 6\times 10^{-14}$ (case 1) and
$10^{-13}$~erg~cm$^{-2}$~s$^{-1}$ (case 2), and $\sim 5\times 10^{-13}$ (case 1) and $3\times
10^{-13}$~erg~cm$^{-2}$~s$^{-1}$ (case 2), in the ranges $0.1-100$~GeV and $>100$~GeV, respectively. We recall that these
fluxes might be easily one order of magnitude higher, since the non-thermal fraction is not much constrained.

\begin{figure}[]
\centering
\includegraphics[width=0.45\textwidth]{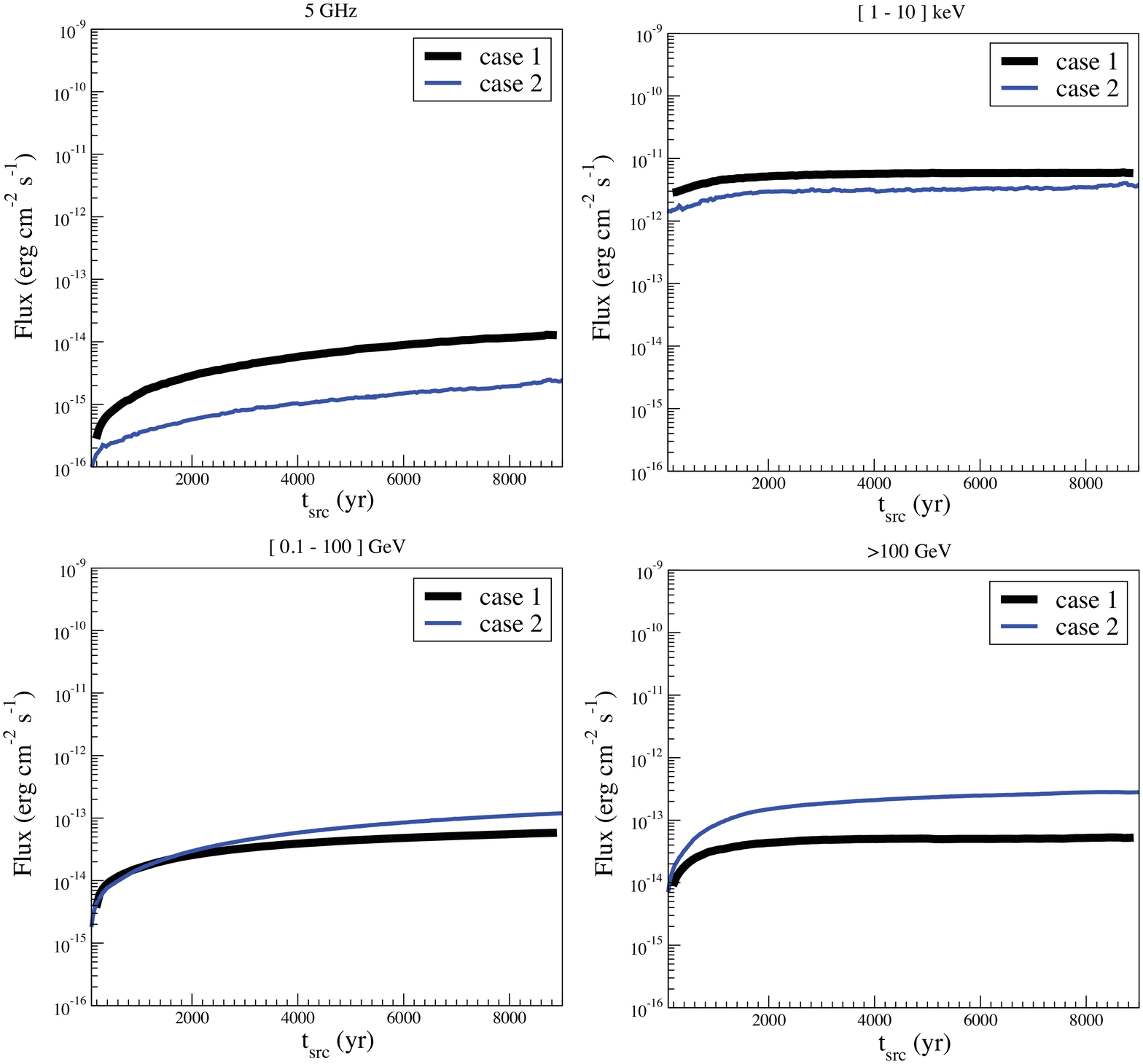}
\caption{Evolution of the 
computed fluxes at 5~GHz (top, left), $1-10$~keV (top, right), $0.1-100$~GeV (bottom, left), and $>100$~GeV (bottom 
right) up to $10^4$~yr, for cases 1 (black/thick lines) and 2 (blue/thin lines).}
\label{fig:lc}
\end{figure}

It is worth noting that the cocoon is in pressure equilibrium with the medium. Since the dynamical timescales of the
SNR/ISM and wind/ISM interaction structures are longer than those generated by the jet, the magnetic field is likely to be
approximately constant with time. In this context, the radio emission grows with the source age due to particle accumulation but the X-ray fluxes do not change
much, because X-ray synchrotron emitting electrons have time to cool down from quite early times. Concerning IC, the
increase of the available cooling time due to longer $t_{\rm src}$ compensates the dilution of the stellar photon field 
(and subsequent reduction of the emission efficiency) 
produced by the increase in cocoon size. This renders fairly constant total IC fluxes, although the spectrum below $\sim 1$~TeV becomes softer with time. In
Fig.~\ref{fig:lc}, we present the evolution of the specific flux at 5~GHz, and the bolometric fluxes between $1-10$~keV,
$0.1-100$~GeV, and $>100$~GeV, up to $10^4$~yr. In general, all the lightcurves show a smooth increase for the considered
times.

From the morphological point of view, the sources discussed here would appear extended at scales of $\sim 10'\,(d/3~{\rm
kpc})^{-1}$ in radio, soft X-rays and high-energy gamma-rays. At hard X-rays and $>1$~TeV, the emission would be rather
concentrated close to the jet reverse shock and possibly the jet recollimation shocks, since the emitting electrons, with
lifetimes $\approx 1.3\times 10^2\,(B/10^{-4}\,{\rm G})^{-2}\,(E/10\,{\rm TeV})^{-1}$~yr, may not have time to fill the
cocoon. 

\section{Discussion}

We briefly discuss here, under the light of our results, the two HMMQs with very clear evidence of jet
interaction with the ISM: SS~433 (case 1) and Cygnus~X-1 (case 2). In the case of SS~433, the jet is strongly interacting
with its supernova remnant (e.g. Zealey et al. 1980). This source is therefore a clear example of a jet/SNR interaction,
but the luminosity of the jet, $\sim 10^{40}$~erg~s$^{-1}$, is much higher than the value adopted here. 
Nevertheless, there
is evidence of non-thermal radio and X-ray emission in the SNR/jet interaction regions, which implies that particle
acceleration is taking place there (e.g. Safi-Harb \& Petre 1999), and possibly gamma-rays, although at present only
upper-limits are available (see Bordas et al. 2010b and references therein). It is worth noting, that the interaction of an
SS~433-like jet with its external medium has been detected even in other, nearby, galaxies (e.g. Pakull et al. 2010). 

Concerning Cygnus~X-1, there is a thin emission region that seems to trace the bow shock formed by the jet in an
inhomogeneous ISM (the counter-jet is not detected; see, e.g., Gallo et al. 2005). A SNR/jet interaction can indeed be
discarded, because the SNR apparently was very weak, as reported by Mirabel \& Rodrigues (2003). The structure is
elongated, and thus a shocked wind/jet origin seems unlikely from observations, although the wind energetics ($\sim
10^{36}-10^{37}$~erg~$^{-1}$) plus medium inhomogeneity might be a possibility, albeit remote. On the other hand, the
deprojected distance of the bow shock to the system is about 10~pc (Gallo et al. 2005), enough for the jet to have already
left the shocked wind/ISM region. As noted above, the non-detection of emission from the cocoon region may be related to jet
disruption. In such a case, radiation may be still generated from particles accelerated via stochastic or shear
acceleration or the jet recollimation shock(s), which is expected even if the jet disrupts at farther distances (see Bordas
et al. 2010a for a similar situation in extragalactic sources). The bow shock in Cygnus X-1 does not show clear signatures
of the system proper motion, which can be explained by the low peculiar velocity of the source with respect to its birth
association, Cygnus~OB3 (Mirabel \& Rodrigues 2003), plus a relatively young jet age (Gallo et al. 2005). 

Other high-mass systems may be good candidates to study the interaction of the jet with its surroundings. Cygnus~X-3, for
instance, has a powerful jet, and radiation has been detected in the surroundings of this source (Heindl et al. 2003;
S\'anchez-Sutil et al. 2008), although the non-thermal nature of this emission and its relation to Cygnus~X-3 are still to
be confirmed. Hints of extended X-ray emission have been also found in LS~I~+61~303 (Paredes et al. 2007; see 
however Rea
et al. 2010), but the microquasar nature of this source is disputed (see, e.g., Bosch-Ramon \& Khangulyan 2009 and
references therein). 

The number of powerful HMMQs interacting with the environment may be of the order of 10 in our Galaxy\footnote{We do not
consider here the impact of microquasars as a class, including low-mass sources, on the ISM properties (see, e.g., Heinz et
al. 2008), but focus on the specific case of jet/environment interaction in HMMQs.} (see, e.g., Perucho et al. 2010). The
small number of sources, and their preferred location in the inner regions of the Galaxy (Bosch-Ramon et al. 2005), makes
their detection difficult due to strong local/foreground absorption of radio and soft X-ray emission. Diffuse hard X-rays,
clustered in the acceleration regions (recollimation and reverse shocks, and possibly the bow shock 
generated in the jet-ISM
interaction), may be better probes to unveil young HMMQs. Very high-energy radiation could be detected from jet/medium
interactions by forthcoming Cherenkov instrumentation (e.g. CTA), which in addition does not suffer of strong gamma-ray
background contamination. It could be however that most HMMQs would be old and fast enough to have a well developed bow
shock driven by the proper motion of the system. In such a case, the jet would be destroyed and mixed with material from
the system motion-driven bow shock, and its radiation, if any, could be hard to detect. 

\section{Summary}

In the present work, we explore different possibilities for the propagation of a HMMQ jet assuming different initial times
for the jet activity to start with respect to the formation of the HMXB: shortly ($40000$~yr) after the SNR explosion of
the progenitor of the compact object; at epochs ($\sim 10^5$~yr) at which only a spherical wind/ISM shock is present; and
at the stages when the system peculiar motion bow-shapes the wind/ISM shock creating strong shocked wind convection {\it
backwards} ($>10^5$~yr). For the two first cases, the jet needs about  $10^4$~yr to reach the unperturbed ISM, a period
much smaller than the potential duration of a HMMQ, which may be up to the lifetime of the remaining massive star, of $\sim
10^6$~yr. For the third case, the jet gets disrupted and mixed with the shocked stellar wind, and advected away from the
stellar bow shock, in just $\sim 10^3$~yr. For the cases in which the jet reaches the normal ISM, its propagation can be
described as in Bordas et al. (2009), although the impact of crossing the shocked wind/SNR ejecta/shocked ISM can lead to
jet disruption, as in FRI jets (e.g. Perucho \& Mart\'i 2007). When the jet is embedded in the hot
shocked wind/SNR ejecta, significant non-thermal emission can be generated in the cocoon region, mainly in radio and
X-rays, but also gamma rays. Hard X-rays and very high-energy photons may be the best probes to unveil the presence of
microquasar jets interacting with their environment.

\begin{acknowledgements}
The  authors thank the referee, Philip Hughes, for his constructive and very encouraging comments.
The research leading to these results has received funding from the European Union
Seventh Framework Programme (FP7/2007-2013) under grant agreement
PIEF-GA-2009-252463
V.B-R. and PB
acknowledge support by the Ministerio de Educaci\'on y Ciencia (Spain) under grant AYA 2007-68034-C03-01, FEDER
funds.  MP acknowledges support from a ``Juan de la Cierva" contract of the Spanish ``Ministerio de Ciencia y
Tecnolog\'{\i}a" and from the Spanish ``Ministerio de Educaci\'on y Ciencia" through grants AYA2007-67627-C03-01,
CSD2007-00050 and AYA2007-67752-C03-02. 
PB acknowledges support from the German Federal Min-
istry of Economics and Technology through DLR grant 50
OG 1001.
PB also acknowledges the excellent work conditions at the INTEGRAL Science Data Center.
The authors acknowledge the {\it ``Servei d'Inform\`atica"} of the {\it ``Universitat de
Val\`encia"} for the  computing time allocated for this project in {\it ``Tirant"}.

\end{acknowledgements}

\end{document}